\documentclass[aps,pre,showkeys,superscriptaddress]{revtex4-1}

\usepackage{graphicx}% Include figure files
\usepackage{amssymb,amsmath,dcolumn}% Align table columns on decimal point
\usepackage{bm}% bold math
\usepackage{color}% color

\usepackage[CJKbookmarks=true,colorlinks,linkcolor=blue,anchorcolor=blue,citecolor=blue,unicode]{hyperref} %
\usepackage{bookmark}

\begin{document}

\preprint{Fractals/Donation behaviors}

\title{Non-Poisson donation behaviors in virtual worlds}

\author{Yan-Hong Yang}
 \affiliation{School of Business, East China University of Science and Technology, Shanghai 200237, China} %
 \affiliation{Department of Physics and Center for Polymer Studies, Boston University, Boston, MA 02215, USA}
 
\author{Ming-Xia Li}
 \affiliation{Research Institute of Sports Economics, East China University of Science and Technology, Shanghai 200237, China} %
 \affiliation{Research Center for Econophysics, East China University of Science and Technology, Shanghai 200237, China} %
 
\author{Wei-Xing Zhou}
 \email{wxzhou@ecust.edu.cn}
 \affiliation{School of Business, East China University of Science and Technology, Shanghai 200237, China} %
 \affiliation{Research Center for Econophysics, East China University of Science and Technology, Shanghai 200237, China} %
 \affiliation{Department of Mathematics, East China University of Science and Technology, Shanghai 200237, China} %

\author{H. Eugene Stanley}
 \affiliation{Department of Physics and Center for Polymer Studies, Boston University, Boston, MA 02215, USA}

\date{\today}

\begin{abstract}
Similar to charitable giving in real world, donation behaviors play an important role in the complex interactions among individuals in virtual worlds. However, it is not clear if the donation process is random or not. We investigate this problem using detailed data from parallel virtual worlds adhered to a massively multiplayer online role-playing game. We find that the inter-donation durations follow power-law-tailed distributions distributed with an average tail exponent close to 1.91, have strong long-range correlations, and possess multifractal features. These findings indicate that the donation process is non-Poissonian, which has potential worth in modeling the complicated individuals behaviors in virtual worlds.
\end{abstract}


\maketitle

\section{Introduction}
\label{S1:Introduction}

Charitable donations which are motivated by altruism and ``warm-glow'' are ubiquitous in real world \cite{OttoniWilhelm-Vesterlund-Xie-2017-AER,Khadjavi-2017-MS}. Intuitively, donation behavior can trigger direct and indirect reciprocity, which is a powerful mechanism for the evolution of cooperation \cite{Nowak-2005-Science}. Cooperation is a key aspect of social evolution, where interactions among individuals affect reproductive success \cite{Nowak-2012-JTB}. Until now, the needed quantities and quality of data related to cooperation in human societies are still difficult to obtain. Alternatively, the online virtual worlds of MMORPGs, where people can learn, work, play and interact with others in a somewhat realistic manner, have great potential for research in behavioral, economic, social and human-centered computer sciences \cite{Bainbridge-2007-Science,Grabowski-Kosinski-2008-APPA}. The availability of big data recorded from massively multiplayer online role-playing games (MMORPGs) enables us to perform quantitative analysis of donation process in virtual worlds.

In recent years, numerous scientific studies have been carried out based on multitudinous data sets collected from online virtual worlds \cite{Grabowski-Kosinski-2008-APPA,Jiang-Zhou-Tan-2009-EPL,Jiang-Ren-Gu-Tan-Zhou-2010-PA,Xie-Li-Jiang-Tan-Podobnik-Zhou-Stanley-2016-SR,Yang-Xie-Li-Jiang-Zhou-2017-CSF,Xie-Yang-Li-Jiang-Zhou-2017-EPJds}. In particular, a pioneering work was done by Edward Castronova, who traveled in a virtual world called ``Norrath'' and performed preliminary analysis of its economy \cite{Castronova-2001-WP}. Studies on the structure and dynamic evolution of social networks in virtual worlds have unveiled intriguing results \cite{Xie-Li-Jiang-Tan-Podobnik-Zhou-Stanley-2016-SR,Xie-Li-Jiang-Zhou-2014-SR,Szell-Lambiotte-Thurner-2010-PNAS,Szell-Thurner-2013-SR,Szell-Thurner-2010-SN}.
Additionally, virtual worlds could establish direct contacts with real social activities, such as marketing \cite{Matsuda-2003-Presence,Castronova-2005-HBR,Hemp-2006-HBR}, and provide opportunities for players to make real money \cite{Papagiannidis-Bourlakis-Li-2008-TFSC}.

Quantitative understanding of regular patterns in human dynamics is of great importance \cite{Zha-Zhou-Zhou-2016-PNAS}. Previously, when detailed and precise records of human activities were rare, individual activities were assumed to follow Poisson processes with exponential distributions of interevent times \cite{Haight-1967}. Recent investigations of interevent intervals between two consecutive interacting actions, such as e-mail communications \cite{Barabasi-2005-Nature,Malmgren-Stouffer-Motter-Amaral-2008-PNAS}, short message correspondences \cite{Wu-Zhou-Xiao-Kurths-Schellnhuber-2010-PNAS}, cell phone conservations \cite{Candia-Gonzalez-Wang-Schoenharl-Madey-Barabasi-2008-JPAMT,Jiang-Xie-Li-Podobnik-Zhou-Stanley-2013-PNAS}, letter correspondences \cite{Oliveira-Barabasi-2005-Nature,Malmgren-Stouffer-Campanharo-Amaral-2009-Science}, online collaborations \cite{Zha-Zhou-Zhou-2016-PNAS}, and order cancelations and equity transactions in financial market \cite{Jiang-Chen-Zhou-2008-PA,Gu-Zhou-2009-EPL,Ni-Jiang-Gu-Ren-Chen-Zhou-2010-PA,Ruan-Zhou-2011-PA,Gu-Xiong-Zhang-Zhang-Zhou-2014-FiP}, indicate that human interactions have non-Poisson characteristics. For instance, Malmgren et al. demonstrated that the approximate power-law scaling of the interevent time distribution of e-mail communication is a consequence of circadian and weekly cycles of human activity. More specially, for the online-offline activities of users in MMORPG, Weibull distribution is unveiled concerning the gaming session durations of users \cite{Jiang-Zhou-Tan-2009-EPL}. In a word, it is meaningful to explore the regular patterns and mechanism of donation process relating to the formation of cooperation among online individuals in virtual worlds.

Complex systems usually exhibit complex behavior characterized by long-term power-law correlations \cite{Sornette-2004}. Extensive studies show that long-term correlations widely exist in many economic or financial, biological, and ecological systems \cite{Gu-Xiong-Zhang-Zhang-Zhou-2014-FiP,Gu-Xiong-Zhang-Chen-Zhang-Zhou-2016-CSF,Peng-Buldyrev-Goldberger-Havlin-Sciortino-Simons-Stanley-1992-Nature,Bunde-Bunde-Havlin-Roman-Goldreich-Schellnhuber-1998-PRL}, especially for human interactive activities in both real and virtual worlds \cite{Jiang-Ren-Gu-Tan-Zhou-2010-PA,Yang-Xie-Li-Jiang-Zhou-2017-CSF,Rybski-Buldyrev-Havlin-Liljeros-Makse-2009-PNAS,Rozenfeld-Rybski-Andrade-Batty-Stanley-Makse-2008-PNAS}.
Specially, long-term correlations become stronger as the human activity level increases \cite{Rybski-Buldyrev-Havlin-Liljeros-Makse-2009-PNAS}. More than ten methods have been
developed to investigate and determine the correlation strength in long-term correlated time series \cite{Taqqu-Teverovsky-Willinger-1995-Fractals,Montanari-Taqqu-Teverovsky-1999-MCM,Delignieres-Ramdani-Lemoine-Torre-Fortes-Ninot-2006-JMPsy,Kantelhardt-2009-ECSS}, including rescaled range (R/S) analysis \cite{Hurst-1951-TASCE}, discrete
wavelet transform \cite{Bunde-Eichner-Havlin-Kantelhardt-2004-PA,Kantelhardt-Roman-Greiner-1995-PA}, wavelet transform module maxima (WTMM) approach \cite{Holschneider-1988-JSP,Muzy-Bacry-Arneodo-1991-PRL,Bacry-Muzy-Arneodo-1993-JSP,Muzy-Bacry-Arneodo-1993-PRE,Muzy-Bacry-Arneodo-1994-IJBC}, the fluctuation analysis (FA) \cite{Peng-Buldyrev-Goldberger-Havlin-Sciortino-Simons-Stanley-1992-Nature}, detrended fluctuation analysis (DFA) \cite{Peng-Buldyrev-Havlin-Simons-Stanley-Goldberger-1994-PRE}, detrending moving average analysis (DMA) \cite{Alessio-Carbone-Castelli-Frappietro-2002-EPJB,Carbone-Castelli-2003-SPIE}, and so on. Scholars have tried to rank the performances of different methods \cite{Xu-Ivanov-Hu-Chen-Carbone-Stanley-2005-PRE,Oswiecimka-Kwapien-Drozdz-2006-PRE,Bashan-Bartsch-Kantelhardt-Havlin-2008-PA,Serinaldi-2010-PA,Jiang-Zhou-2011-PRE,Huang-Schmitt-Hermand-Gagne-Lu-Liu-2011-PRE,Bryce-Sprague-2012-SR,Shao-Gu-Jiang-Zhou-2015-Fractals,Grech-Mazur-2013-PRE,Grech-Mazur-2015-APPA} and not reached clear-cut consensus. It is not unreasonable that the conclusions are mixed because different studies used different time series generators and different lengths. However, there is wide agreement that DMA and DFA are ``The Methods of Choice'' in determining the long memory strength of time series \cite{Shao-Gu-Jiang-Zhou-Sornette-2012-SR}. Therefore, we apply the DMA and DFA methods to investigate the memory effect of inter-donation duration series in this paper.

Likewise, multifractals are ubiquitous in natural and social sciences \cite{Mandelbrot-1983,Jiang-Xie-Zhou-Sornette-2018-XXX}. A wealth of methods have been invented to characterize the hidden multifractal nature of different social variables, such as the structure function method \cite{Kolmogorov-1962-JFM,VanAtta-Chen-1970-JFM,Anselmet-Gagne-Hopfinger-Antonia-1984-JFM,Ghashghaie-Breymann-Peinke-Talkner-Dodge-1996-Nature},
the partition function method
 \cite{Grassberger-1983-PLA,Hentschel-Procaccia-1983-PD,Grassberger-1985-PLA,Halsey-Jensen-Kadanoff-Procaccia-Shraiman-1986-PRA,Xie-Jiang-Gu-Xiong-Zhou-2015-NJP}, the multiplier method \cite{Chhabra-Sreenivasan-1992-PRL,Jouault-Lipa-Greiner-1999-PRE,Jiang-Zhou-2007-PA}, the wavelet transform approaches \cite{Muzy-Bacry-Arneodo-1991-PRL,Muzy-Bacry-Arneodo-1993-PRE}, the multifractal detrending moving average analysis (MFDMA) \cite{Gu-Zhou-2010-PRE}, the multifractal detrended fluctuation analysis (MFDFA) \cite{CastroESilva-Moreira-1997-PA,Weber-Talkner-2001-JGR,Kantelhardt-Zschiegner-KoscielnyBunde-Havlin-Bunde-Stanley-2002-PA},
 and so forth. In addition, Gu et al. found that the backward MFDMA exhibits the best performance when compared with the centered MFDMA, forward MFDMA, and MFDFA \cite{Gu-Zhou-2010-PRE}. We hence adopt the backward MFDMA method in this work.

The rest of this paper is organized as follows. Section~\ref{S1:Data description} describes the data used in our study, including some basic statistical properties of user donation actions and inter-donation durations. Section \ref{S1:PDF} studies the probability distribution of inter-donation durations and Section~\ref{S1:LRC} estimates its memory behavior using advanced statistical methods. We further investigate multifractal nature of inter-donation durations in Section~\ref{S1:Multifractal}. Finally, We summarize our findings in Section~\ref{S1:Conclusion}.

\section{Data description}
\label{S1:Data description}

MMORPG is the typical online virtual worlds with sophisticated interactions. Our study is based on a huge database recorded from 98 servers of a popular MMORPG in China, from 16 May 2011 to 31 December 2012, to uncover the patterns characterizing online virtual worlds. The dataset contains all in-game action logs for more than 120 days in each server, in which the time span varies with the loss proportion of characters and time stamps are accurate to 1 second. Nevertheless, we mainly focus on item logs in this study. A donation action is written to the log file when individuals give items to or receive from others without exchanging any money or items. Therefore, the donation actions in a log file are arranged according to an increasing order of donation moments. Each donation contains at least three pieces of information: the donor ID, the donee ID, and the corresponding donation time. For each individual in MMORPG, we collect all the associated donations. During this period, on average, there were more than 100000 characters who logged on the game for each server and the majority of them had made donation. For security sake, the true characters IDs have been encrypted into numbers from 1 to ordinal number of the last ID for each virtual world.

\begin{table}[htp]
  \setlength\tabcolsep{2.5pt}
  \centering
  \caption{Descriptive statistics of inter-donation durations for 98 virtual worlds. $N$ is the number of donations, $\rho$ is the ratio of the number of simultaneously happening donations $N^0$ to $N$, $\langle{\tau}\rangle$ is the average inter-donation duration in units of second.}
  \label{TB:Statistics}
  \small
   \begin{tabular}{*{9}{c}}
   \hline\hline
  Code&$N$ &$\rho$ & $\langle \tau \rangle$ && Code&$N$ &$\rho$ & $\langle \tau \rangle$\\
   \hline
 0001&1705228 &0.617 & 20.50 && 0050&1535839 &0.627 & 24.47\\
 0002&2239240 &0.608 & 21.85 && 0051&1601163 &0.613 & 27.54\\
 0003&3296855 &0.611 & 14.70 && 0052&2573136 &0.578 & 15.73\\
 0004&1532055 &0.630 & 32.51 && 0053&1356094 &0.606 & 31.95\\
 0005&1334578 &0.634 & 37.71 && 0054&1131281 &0.592 & 36.76\\
 0006&2444375 &0.614 & 19.26 && 0055&1216377 &0.603 & 35.09\\
 0007&1660899 &0.636 & 28.69 && 0056&1689275 &0.608 & 25.75\\
 0008&2320411 &0.614 & 19.37 && 0057&2605908 &0.597 & 16.20\\
 0009&1998936 &0.635 & 24.55 && 0058&1567798 &0.579 & 25.76\\
 0010&1275256 &0.595 & 26.66 && 0059&1789417 &0.611 & 22.66\\
 0011&1082580 &0.590 & 31.00 && 0060&847119 &0.614 & 40.77\\
 0012&1094432 &0.630 & 45.74 && 0061&1324435 &0.594 & 29.20\\
 0013&554977 &0.602 & 46.06 && 0062&2052950 &0.623 & 20.35\\
 0014&1173571 &0.601 & 28.49 && 0063&1299678 &0.602 & 30.52\\
 0015&1079154 &0.598 & 30.71 && 0064&1096244 &0.591 & 35.15\\
 0016&1238259 &0.610 & 27.58 && 0065&1055676 &0.608 & 37.95\\
 0017&886742 &0.596 & 37.73 && 0066&797059 &0.609 & 50.36\\
 0018&1139159 &0.611 & 30.33 && 0067&1221146 &0.624 & 34.27\\
 0019&1075634 &0.585 & 29.51 && 0068&2109187 &0.608 & 19.14\\
 0020&1282801 &0.589 & 25.60 && 0069&1222507 &0.605 & 37.53\\
 0021&3052685 &0.612 & 20.52 && 0070&1650616 &0.597 & 23.92\\
 0022&3178363 &0.597 & 19.00 && 0071&1814959 &0.609 & 22.36\\
 0023&1694692 &0.614 & 21.28 && 0072&1773985 &0.598 & 22.17\\
 0024&1494519 &0.617 & 24.31 && 0073&1293745 &0.597 & 30.38\\
 0025&1148468 &0.627 & 32.54 && 0074&1219383 &0.552 & 28.94\\
 0026&1485269 &0.638 & 25.87 && 0075&900405 &0.597 & 43.73\\
 0027&1248578 &0.633 & 29.63 && 0076&1757164 &0.598 & 22.46\\
 0028&1090820 &0.632 & 33.81 && 0077&1195309 &0.594 & 33.65\\
 0029&1377820 &0.646 & 25.39 && 0078&1650125 &0.620 & 26.17\\
 0030&910070 &0.652 & 41.05 && 0079&1731510 &0.593 & 23.30\\
 0031&3283140 &0.606 & 16.14 && 0080&1404610 &0.596 & 28.88\\
 0032&2249825 &0.606 & 23.54 && 0081&3028075 &0.576 & 12.85\\
 0033&1735769 &0.574 & 26.53 && 0082&1423213 &0.593 & 28.44\\
 0034&2359477 &0.572 & 19.43 && 0083&1217974 &0.577 & 32.01\\
 0035&2735192 &0.610 & 16.54 && 0084&1366396 &0.590 & 29.41\\
 0036&2306332 &0.621 & 20.24 && 0085&1148700 &0.592 & 35.18\\
 0037&3349221 &0.618 & 13.74 && 0086&739746 &0.566 & 50.92\\
 0038&1966960 &0.602 & 22.47 && 0087&962872 &0.591 & 41.77\\
 0039&1410212 &0.599 & 31.29 && 0088&1149438 &0.593 & 35.18\\
 0040&1504820 &0.592 & 28.82 && 0089&1319263 &0.595 & 36.32\\
 0041&1477299 &0.580 & 28.46 && 0090&1958365 &0.619 & 17.24\\
 0042&1512435 &0.607 & 29.75 && 0091&1382076 &0.625 & 24.90\\
 0043&844896 &0.634 & 35.87 && 0092&1059380 &0.616 & 31.58\\
 0044&1924209 &0.644 & 20.41 && 0093&1096790 &0.621 & 30.81\\
 0045&1180294 &0.628 & 31.93 && 0094&1166467 &0.616 & 28.84\\
 0046&686874 &0.610 & 38.64 && 0095&1251507 &0.617 & 26.91\\
 0047&516067 &0.613 & 49.56 && 0096&1741728 &0.647 & 29.19\\
 0048&977913 &0.612 & 36.90 && 0097&1370604 &0.641 & 33.91\\
 0049&915043 &0.630 & 32.73 && 0098&1908909 &0.626 & 27.75\\
   \hline\hline
   \end{tabular}
\end{table}

\begin{figure}[!htb]
  \centering
  \includegraphics[width=\linewidth]{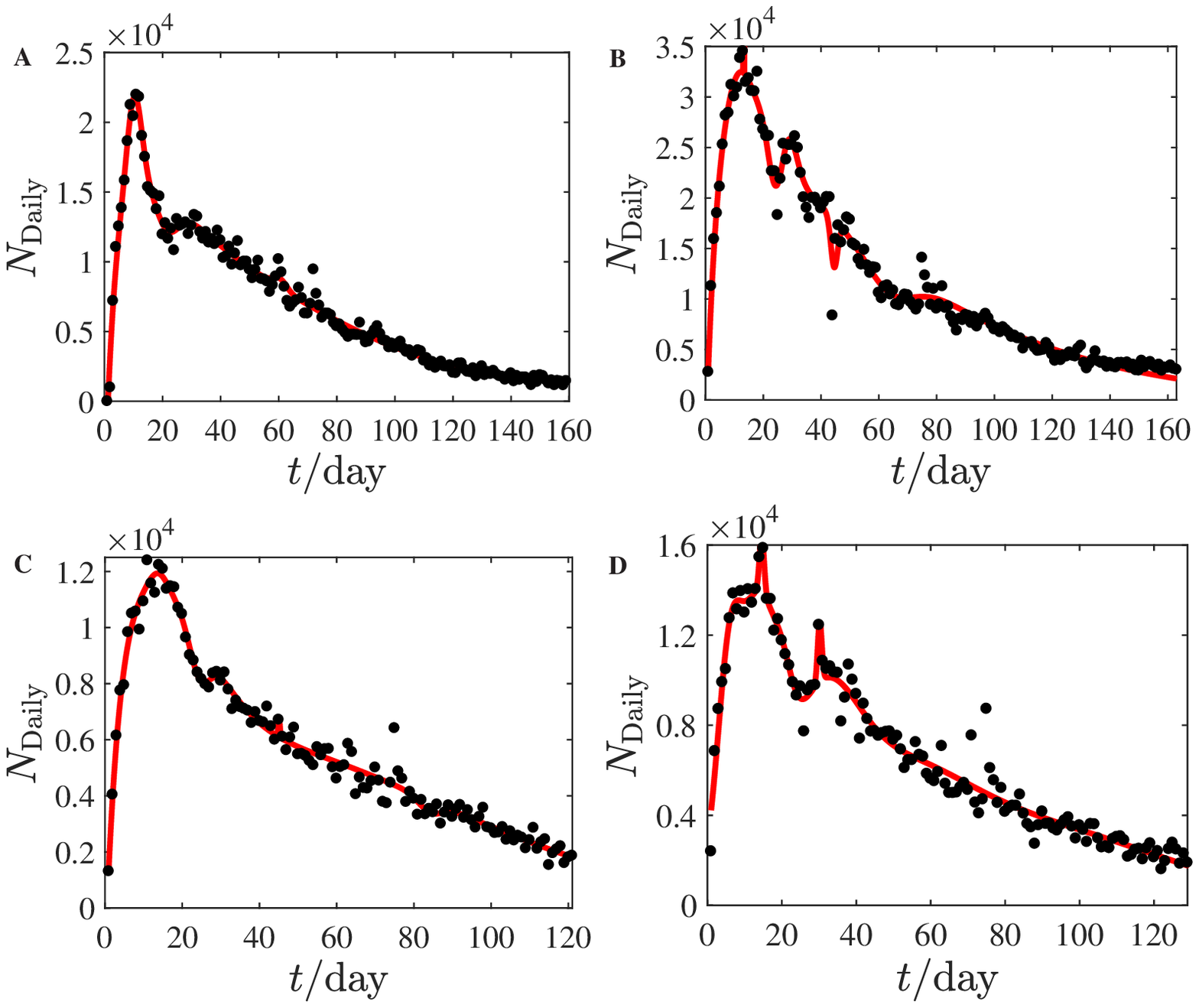}
  \caption{Daily evolution of donation actions in four different virtual worlds 0028, 0044, 0046 and 0049, which correspond to (A-D), respectively. The dots are the daily donation actions. The continuous curves are the polynomial fits to the data. Other virtual worlds have very similar patterns.}
  \label{Fig:XCB:IDD:Evolution}
\end{figure}

We obtain the entire donation number $N$ for each virtual world, which is listed in Table~\ref{TB:Statistics}. We find that the donation number $N$ fluctuates within a wide range, from 516067 to 3349221. Meanwhile, we have removed the abnormal donations (e.g. when the servers were scheduled for maintaining or during game version updating) of the 98 virtual worlds in order to ensure statistical significance. In addition, Fig.~\ref{Fig:XCB:IDD:Evolution} presents the daily evolution of donation actions in four different virtual worlds, which exhibits the similar shape with the evolution of the number of active characters \cite{Xie-Li-Jiang-Tan-Podobnik-Zhou-Stanley-2016-SR,Yang-Xie-Li-Jiang-Zhou-2017-CSF}.
With the development of a virtual world, the number of daily donations $N_{\rm{Daily}}$ increases and reaches a maximum around the 10th day and then decays. Especially, there exists another evident local hump around the 30th day, which is mainly caused by some new marketing actions organized by the online game operators. Additionally, We find that curves of the rest daily donation series of the 98 virtual worlds almost share the same shapes as in Fig.~\ref{Fig:XCB:IDD:Evolution} except for some special dates.

%
%\begin{figure}[!htp]
%  \centering
%  \includegraphics[width=7cm]{Fig_XCB_IDD_Arrow.eps}\\
%  \caption{Schematic chart of donation logs for individuals in MMORPG.}
%  \label{Fig:xcb:idd:arrow}
%\end{figure}

In the paper, the inter-donation duration is defined as the interval between two consecutive donations in units of second, which reads
\begin{equation}
  \tau_i=t_{i+1}^{d}-t_i^{d}~,
  \label{Eq:df_cd}
\end{equation}
where $t_i^{d}$ is the time when $i$-th donation takes place. Although the time resolution of our data is as precise as 1 second, on average, there are still more than 55\% donations stamped with the same time which are presented in Table~\ref{TB:Statistics} using $\rho$, indicating that the inter-donation duration is vanishing between the two corresponding donations. For convenience, we treat the donations occurring at the same time as one donation at that time. Therefore, vanishing durations are excluded. For all the 98 virtual worlds, we calculate the average values $\langle{\tau}\rangle$ of inter-donation duration series which vary from 12.85 seconds to 50.92 seconds, and depict the results in Table~\ref{TB:Statistics}.

\section{Probability distributions of inter-donation durations}
\label{S1:PDF}

The probability distribution of a random variable is of essential importance since it can fully determine the moments of the variable and may has a direct relationship to the memory effects and multifractality of the time series \cite{Gu-Xiong-Zhang-Chen-Zhang-Zhou-2016-CSF,Zhou-2012-CSF}. In this section, we focus on investigating the probability distributions of inter-donation durations $\tau$ of all the 98 virtual worlds.

\begin{figure}[htp]
  \centering
   \includegraphics[width=\linewidth]{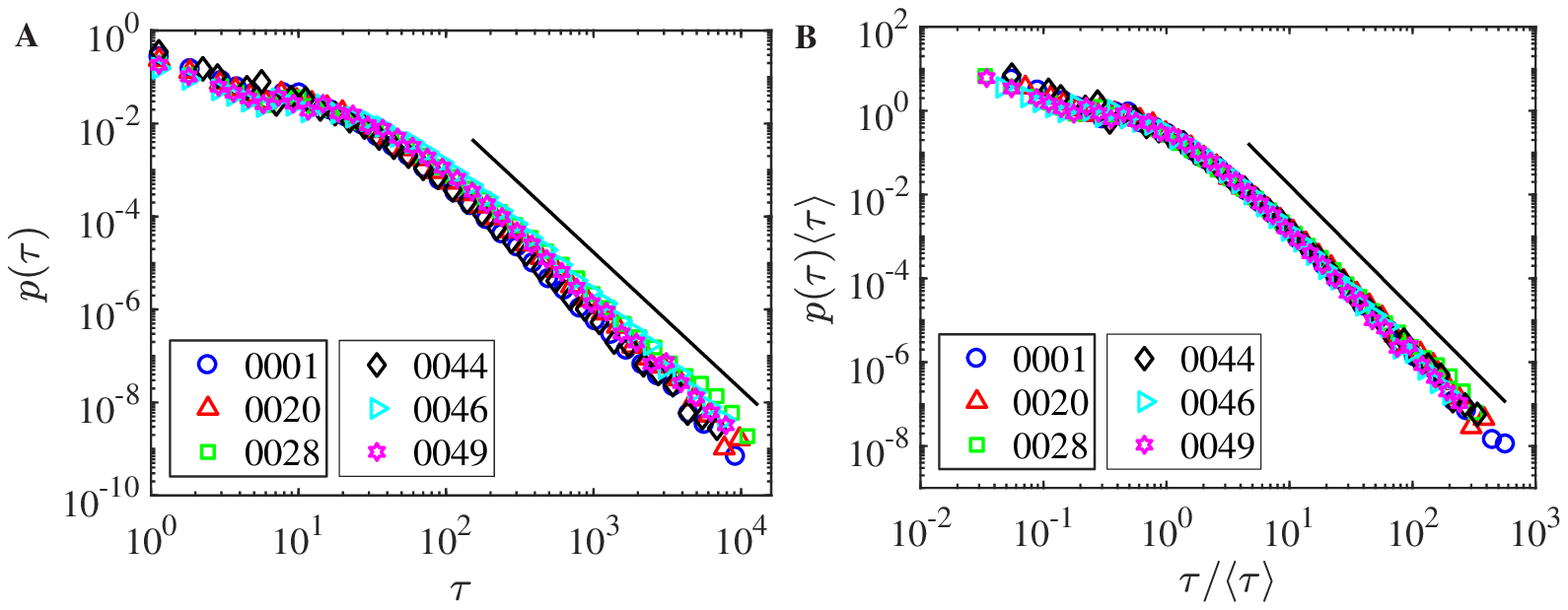}
  \caption{Probability distributions (A) and rescaled probability distributions (B) of inter-donation durations $\tau$ of six typical virtual worlds 0001, 0020, 0028, 0044, 0046 and 0049. The slopes of the solid lines are -2.95.}
  \label{Fig:XCB:IDD:PDF}
\end{figure}

The associated empirical probability distributions $p(\tau)$ of six randomly chosen virtual worlds are presented in Fig.~\ref{Fig:XCB:IDD:PDF}(A), which exhibit a clear scaling decay. In addition, we rescale the inter-donation duration $\tau$ to $\tau/\langle{\tau}\rangle$ and the probability distributions $p(\tau)$ to $p(\tau)\langle{\tau}\rangle$. The rescaled probability distributions of inter-donation durations for the same six virtual worlds are presented in Fig.~\ref{Fig:XCB:IDD:PDF}(B), which almost collapse together and show a perfect scaling behavior at tails. Consequently, we conjecture that these distributions have power-law tails:
\begin{equation}
 p(\tau) \sim \tau^{-(\beta+1)}~~~{\rm{for}}~~\tau \geq \tau_{\min},
 \label{Eq:tau:PL}
\end{equation}
where $\beta$ is the power-law exponent and $\tau_{\min}$ is the lower threshold of the scaling range of the power-law decay.

\begin{table*}[htp]
  \setlength\tabcolsep{8pt}
  \caption{{Characteristic parameters in the power-law distributions of inter-donation duration $\tau$ for the 98 virtual worlds based on the Kolmogorov-Smirnov tests and the maximum likelihood estimation.}}
   \label{TB:PL:Parameters}
  \small
  \centering
   \begin{tabular}{*{13}{c}}
   \hline\hline
  Code&$\tau_{\min}$ &$\beta$ & $\sigma_{\beta}$ &$KS$&$p$-value&&Code&$\tau_{\min}$ &$\beta$ & $\sigma_{\beta}$ &$KS$&$p$-value \\
   \hline
 0001&147 &1.95 &0.020 &0.007 &0.57 &&0050&138 &1.99 &0.018 &0.008 &0.26 \\
 0002&118 &1.78 &0.012 &0.011 &0.01 &&0051&215 &1.63 &0.016 &0.012 &0.02 \\
 0003&107 &2.09 &0.018 &0.005 &0.56 &&0052&128 &1.77 &0.015 &0.006 &0.48 \\
 0004&228 &1.68 &0.016 &0.007 &0.53 &&0053&142 &1.53 &0.011 &0.016 &0.00 \\
 0005&633 &1.90 &0.035 &0.016 &0.25 &&0054&367 &1.69 &0.022 &0.016 &0.06 \\
 0006&119 &1.78 &0.013 &0.004 &0.67 &&0055&232 &1.72 &0.017 &0.012 &0.04 \\
 0007&269 &1.74 &0.021 &0.010 &0.35 &&0056&398 &2.05 &0.041 &0.010 &0.92 \\
 0008&448 &2.14 &0.049 &0.010 &0.99 &&0057&175 &2.23 &0.029 &0.011 &0.21 \\
 0009&375 &1.91 &0.033 &0.010 &0.68 &&0058&189 &1.76 &0.018 &0.007 &0.53 \\
 0010&143 &1.93 &0.016 &0.011 &0.02 &&0059&171 &1.90 &0.020 &0.005 &0.86 \\
 0011&232 &1.93 &0.025 &0.010 &0.26 &&0060&131 &1.85 &0.014 &0.014 &0.00 \\
 0012&386 &1.80 &0.023 &0.016 &0.02 &&0061&248 &1.74 &0.020 &0.008 &0.51 \\
 0013&114 &1.68 &0.013 &0.007 &0.16 &&0062&246 &2.27 &0.041 &0.009 &0.87 \\
 0014&267 &1.95 &0.029 &0.006 &0.98 &&0063&216 &1.80 &0.020 &0.004 &0.99 \\
 0015&256 &1.87 &0.025 &0.009 &0.60 &&0064&281 &1.86 &0.025 &0.006 &0.93 \\
 0016&367 &1.97 &0.038 &0.010 &0.86 &&0065&517 &1.87 &0.038 &0.010 &0.84 \\
 0017&311 &1.74 &0.024 &0.007 &0.94 &&0066&388 &1.81 &0.025 &0.016 &0.03 \\
 0018&330 &1.90 &0.034 &0.006 &1.00 &&0067&177 &1.75 &0.016 &0.007 &0.34 \\
 0019&207 &1.81 &0.021 &0.009 &0.45 &&0068&177 &2.24 &0.028 &0.010 &0.30 \\
 0020&445 &1.97 &0.048 &0.009 &1.00 &&0069&115 &1.79 &0.010 &0.008 &0.00 \\
 0021&122 &1.89 &0.012 &0.009 &0.01 &&0070&211 &2.06 &0.027 &0.009 &0.43 \\
 0022&127 &2.02 &0.015 &0.010 &0.01 &&0071&144 &1.89 &0.018 &0.006 &0.59 \\
 0023&128 &2.10 &0.020 &0.004 &0.91 &&0072&138 &1.92 &0.018 &0.005 &0.82 \\
 0024&196 &2.11 &0.028 &0.005 &1.00 &&0073&193 &1.83 &0.019 &0.006 &0.77 \\
 0025&287 &1.94 &0.030 &0.009 &0.79 &&0074&238 &1.75 &0.020 &0.007 &0.70 \\
 0026&158 &2.06 &0.021 &0.007 &0.51 &&0075&460 &1.79 &0.031 &0.011 &0.59 \\
 0027&181 &2.01 &0.024 &0.008 &0.54 &&0076&161 &1.91 &0.020 &0.006 &0.68 \\
 0028&321 &1.86 &0.031 &0.008 &0.92 &&0077&613 &1.81 &0.039 &0.013 &0.80 \\
 0029&153 &1.88 &0.019 &0.005 &0.82 &&0078&303 &1.97 &0.030 &0.006 &0.99 \\
 0030&751 &2.17 &0.071 &0.018 &0.79 &&0079&333 &2.34 &0.053 &0.011 &0.89 \\
 0031&96 &1.88 &0.012 &0.005 &0.24 &&0080&328 &1.99 &0.031 &0.010 &0.67 \\
 0032&265 &1.82 &0.020 &0.009 &0.28 &&0081&139 &2.27 &0.029 &0.007 &0.80 \\
 0033&271 &1.72 &0.019 &0.010 &0.18 &&0082&161 &1.83 &0.015 &0.010 &0.03 \\
 0034&214 &1.68 &0.016 &0.006 &0.58 &&0083&156 &1.79 &0.015 &0.006 &0.34 \\
 0035&155 &1.93 &0.020 &0.008 &0.38 &&0084&215 &1.77 &0.018 &0.008 &0.30 \\
 0036&172 &1.78 &0.017 &0.005 &0.95 &&0085&296 &1.87 &0.023 &0.011 &0.28 \\
 0037&113 &2.37 &0.025 &0.006 &0.68 &&0086&423 &1.68 &0.025 &0.014 &0.10 \\
 0038&131 &2.05 &0.016 &0.012 &0.00 &&0087&347 &1.76 &0.024 &0.005 &1.00 \\
 0039&249 &1.85 &0.023 &0.007 &0.61 &&0088&304 &1.82 &0.025 &0.005 &0.98 \\
 0040&190 &1.79 &0.016 &0.011 &0.04 &&0089&102 &1.82 &0.009 &0.005 &0.18 \\
 0041&261 &1.92 &0.025 &0.010 &0.22 &&0090&103 &2.04 &0.019 &0.004 &0.88 \\
 0042&211 &1.82 &0.019 &0.004 &0.97 &&0091&153 &1.89 &0.020 &0.006 &0.68 \\
 0043&376 &1.95 &0.040 &0.007 &1.00 &&0092&230 &1.92 &0.025 &0.008 &0.72 \\
 0044&147 &1.93 &0.020 &0.004 &0.97 &&0093&167 &1.93 &0.021 &0.008 &0.31 \\
 0045&253 &1.96 &0.026 &0.011 &0.31 &&0094&208 &2.04 &0.026 &0.007 &0.89 \\
 0046&222 &1.92 &0.027 &0.008 &0.75 &&0095&136 &1.94 &0.020 &0.006 &0.48 \\
 0047&161 &1.75 &0.017 &0.009 &0.14 &&0096&152 &1.84 &0.014 &0.006 &0.50 \\
 0048&176 &1.78 &0.017 &0.008 &0.22 &&0097&276 &1.71 &0.020 &0.005 &0.99 \\
 0049&180 &1.95 &0.024 &0.007 &0.72 &&0098&140 &1.76 &0.012 &0.006 &0.22 \\
   \hline\hline
   \end{tabular}
\end{table*}

In order to capture the tail behavior of the distribution, we need to conduct an objective analysis. Based on  maximum likelihood estimation method (MLE) and the  Kolmogorov-Smirnov statistic ($KS$), Clauset et al. proposed an efficient quantitative method to test if the tail has a power-law form and, if so, to estimate the power-law exponent $\beta$ for the data greater than or equal to a lower bound $\tau_{\min}$ \cite{Clauset-Shalizi-Newman-2009-SIAMR}. Because the values of $\tau$ are positive integers, we only focus on the discrete case of this method. The $KS$ statistic is defined as
\begin{equation}
  KS=\max_{\tau \geq \tau_{\min}}(|P-F_{\rm{PL}}|),
  \label{Eq:KS}
\end{equation}
where $P$ is the cumulative distribution of inter-donation durations $\tau$ and $F_{\rm{PL}}$ is the cumulative distribution of the best power-law fit. The threshold $\tau_{\min}$ is determined by minimizing the $KS$ statistic. Then the power-law exponent $\beta$ of the data in the range $\tau \geq \tau_{\min}$ can be estimated using the MLE, that is,
\begin{equation}\label{Eq:Discrete:Beta}
  \beta \simeq m\left[\sum_{i=1}^{m}\mathrm{ln}\frac{\tau_{i}}{|\tau_{\min}-\frac{1}{2}}\right]^{-1}.
\end{equation}
where $m$ is the number of the data points in the range $\tau>\tau_{\min}$. The standard error $\sigma_{\beta}$ on the power-law exponent $\beta$ is derived from a quadratic approximation to the log-likelihood at its maximum, which reads
\begin{equation}\label{Eq:Discrete:Sigma}
  \sigma_{\beta}=\frac{1}{\sqrt{m\left[\frac{\zeta''(\beta+1,\tau_{\min})}{\zeta(\beta+1,\tau_{\min})}
  -\left[\frac{\zeta'(\beta+1,\tau_{\min})}{\zeta(\beta+1,\tau_{\min})}\right]^{2}\right]}},
\end{equation}
and $\zeta$ is the generalized or Hurwitz zeta function.

Following Clauset et al. \cite{Clauset-Shalizi-Newman-2009-SIAMR}, we conduct the bootstrap test to check whether the power-law tail is a plausible fit to the inter-donation durations $\tau$. In doing so, we generate 2500 realizations of synthetic data for each distribution, for which we wish the $p$-value to be accurate to about 2 decimal digits \cite{Clauset-Shalizi-Newman-2009-SIAMR}. For each realization, we fit synthetic data set individually to its own power-law model and calculate the statistic ${KS}_{\mathrm{sim}}$ for each realization relative to its own model, which is as follows:
 \begin{equation}\label{Eq:KS:Test:sim}
  {KS}_{\mathrm{sim}}=\max(|P_{\mathrm{sim}}-F_{\mathrm{PL}}|),
\end{equation}
where $P_{\mathrm{sim}}$ is the cumulative distribution of the synthetic realization. We thus obtain the $p$-value:
\begin{equation}
  p{\textrm{-value}} = \frac{\#(KS_{\rm{sim}}>KS)}{2500},
\end{equation}
where the numerator is the number of realizations with $KS_{\rm{sim}}>KS$. The meaning of this test is that the investigated inter-donation durations have the power-law tails with a probability of $p$. The resulting $KS$ values and the corresponding $p$-values are given in Table~\ref{TB:PL:Parameters}. Applying this approach \cite{Clauset-Shalizi-Newman-2009-SIAMR}, we identify 84 cases out of the 98 inter-donation duration series that have power-law tails, in which the $p$-values are greater than 5\%. Meanwhile, the determined  characteristic parameters $\tau_{\min}$, $\beta$ and $\sigma_{\beta}$ are presented in Table~\ref{TB:PL:Parameters}. We find that the power-law tail exponent $\beta$ mainly concentrates in the range $[1.7,2.1]$ and combines with a mean close to 1.91, while the lower bound $\tau_{\min}$ mainly concentrates in the range $[100,300]$ and combines with an average value close to 244. Furthermore, for the group of 84 inter-donation duration series with power-law tails ($p$-values greater than 5\%), Table~\ref{TB:PL:Parameters} shows that there are only 6 $p$-values less than 0.2 and 78\% of $p$-values are greater than 0.3. In a word, this test confirms that most virtual worlds have power-law tails in the inter-donation duration distributions, thus the donations of individuals cannot be described by the Poisson process.

\section{Long-range correlation}
\label{S1:LRC}

In this section, we adopt the DMA and DFA methods to investigate if there are long-range corrections in inter-donation duration time series. As mentioned, DMA and DFA are among the most effective and the most extensively used methods \cite{Shao-Gu-Jiang-Zhou-Sornette-2012-SR}. The procedures of the DMA and DFA
methods are briefly described below, which share the same framework \cite{Gu-Xiong-Zhang-Chen-Zhang-Zhou-2016-CSF,Jiang-Xie-Zhou-Sornette-2018-XXX}.

For a given inter-donation duration series $\{\tau_i|i = 1, 2, ..., N\}$, we calculate the cumulative summation series $y_i$ as follows,
\begin{equation}
 y_i=\sum_{j=1}^{i}(\tau_{j}-\langle\tau\rangle),~~~~i=1,2,...,N. \label{Eq:yi}
\end{equation}
where $\langle{\tau}\rangle$ is the sample mean of the $\tau_i$ series. The series $y_i$ is covered by $N_s$ disjoint boxes with the same size $s$. When the whole series $y_i$ cannot be completely covered by $N_s$ boxes, we can utilize $2N_s$ boxes to cover the series from both ends of the series. In each box, a trend function ${\tilde{y}_i}$ of the sub-series is determined. The residuals are calculated by
\begin{equation}
  \epsilon(i) = y_i-{\tilde{y}_i}.
  \label{Eq:epsilon}
\end{equation}

The main difference between the DFA and DMA algorithms is the determination of the ``local trend function'' $\widetilde{y}_i$, which is dependent of the box size $s$. The local trend $\widetilde{y}_i$ could be polynomials, which recovers the DFA method \cite{Peng-Buldyrev-Havlin-Simons-Stanley-Goldberger-1994-PRE}. In the DMA approach, one calculates the moving average function $\widetilde{y}_i$ in a moving window \cite{Arianos-Carbone-2007-PA},
\begin{equation}
  \widetilde{y}_i(s)=\frac{1}{s}\sum_{k=-\lfloor(s-1)\theta\rfloor}^{\lceil(s-1)(1-\theta)\rceil}y_{i-k},
  \label{Eq:1ddma:y1}
\end{equation}
where $\theta$ is the position parameter with the value varying in the range $[0,1]$. Specially, the cases $\theta=0$, $\theta=0.5$ and  $\theta=1$ respectively correspond to the backward detrending moving average (BDMA) method, the centred detrending moving average (CDMA) method and the forward detrending moving average (FDMA) method.
The local fluctuation function $F_v(s)$ in the $v$-th box is defined as the root-mean-square of the residuals:
\begin{equation}
  \left[F_v(s)\right]^2 = \frac{1}{s}\sum_{i=(v-1)s+1}^{vs} \left[\epsilon(i)\right]^2~.
  \label{Eq:fv:s}
\end{equation}
The overall fluctuation function is calculated as follows:
\begin{equation}
  F(s) = \left\{\frac{1}{N_s}\sum_{v=1}^{N_s} {F_v^2(s)}\right\}^{\frac{1}{2}},
  \label{Eq:F2s}
\end{equation}
For most time series with fractal nature, one has:
\begin{equation}
 F(s) \sim s^{H},
 \label{Eq:H}
\end{equation}
where $H$ can be roughly viewed as the Hurst exponent (To be precise, $H$ signifies the DFA or DMA scaling exponent). Hence, if $H$ is significantly greater than 0.5, the inter-donation duration series $\tau$ is positively correlated. If $H$ is insignificantly different from 0.5, the duration series $\tau$ is uncorrelated. If $H$ is significantly smaller than 0.5, the duration series $\tau$ is negatively correlated. When $H$ is compared with 0.5, statistical tests are necessary \cite{Jiang-Xie-Zhou-2014-PA}.

\begin{figure}[htb]
  \centering
  \includegraphics[width=\linewidth]{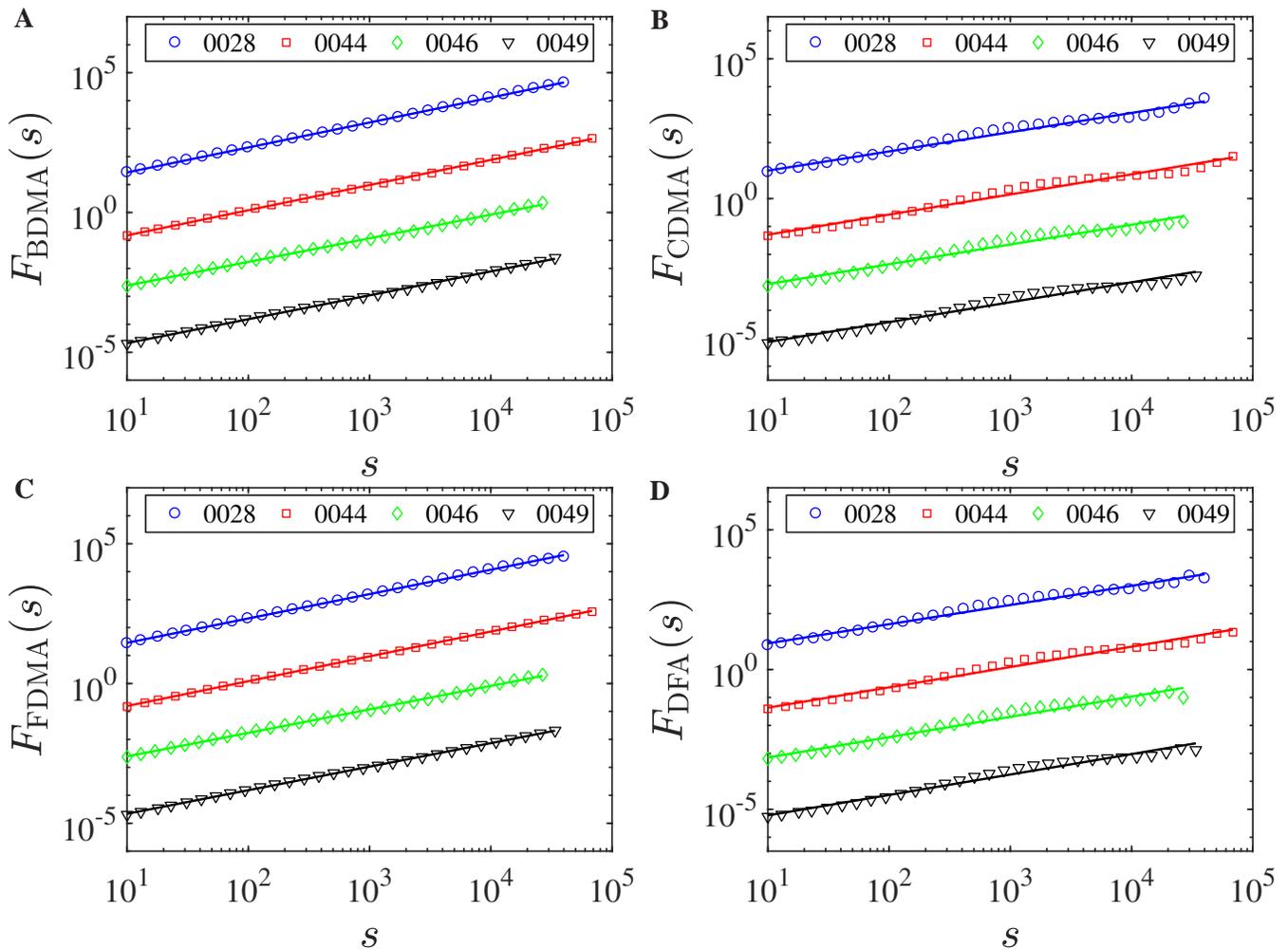}
  \caption{Plots of the fluctuation functions $F(s)$ with respect to the box size $s$ for four inter-donations duration series. The solid lines are power-law fits to the empirical data. The curves have been shifted vertically for clarity. (A) DMA with $\theta = 0$. (B) DMA with $\theta = 0.5$. (C) DMA with $\theta = 1$. (D) DFA.}
  \label{Fig:XCB:IDD:Fs}
\end{figure}

Fig.~\ref{Fig:XCB:IDD:Fs} illustrates the detrended fluctuation function $F(s)$ with respect to the box size $s$ using DMA ($\theta = 0$ in panel A, $\theta = 0.5$ in panel B, and $\theta = 1$ in panel C) and DFA (panel D) for four typical virtual worlds 0028, 0044, 0046 and 0049. Each curve reveals excellent power-law scaling over more than three orders of magnitude. By observing Fig.~\ref{Fig:XCB:IDD:Fs}, it's worth noting that the power-law scaling detected by CDMA and DFA shows slightly worse performance comparing with BDMA and FDMA, which has minor influence on detecting long-range correlations of $\tau$ (Table~\ref{TB:DFA:DMA:H}). The DMA or DFA scaling exponents $H$ of 98 inter-donation duration series are estimated according to the power-law relation $F_{\rm{DMA}}(s)\sim s^{H^{\rm{DMA}}}$ or $F_{\rm{DFA}}(s)\sim s^{H^{\rm{DFA}}}$, which are the slopes of solid lines shown in the log-log plot of Fig.~\ref{Fig:XCB:IDD:Fs}. We calculate the scaling exponents $H$ for all 98 virtual worlds in Table~\ref{TB:DFA:DMA:H}.

\begin{table*}[htb]
%\addtolength{\tabcolsep}{2pt}
  \caption{The scaling exponents qualifying long-range correlations of inter-donation durations $\tau$ for the 98 virtual worlds based on the DFA and DMA methods. $\overline{H}_{\rm{SFL}}$ is the average scaling exponent of 100 shuffled inter-donation durations series. BDMA, CDMA and FDMA are respectively backward, centred and forward DMA.}
   \label{TB:DFA:DMA:H}
\resizebox{\textwidth}{!}{
\small
\centering
   \begin{tabular}{*{19}{c}}
   \hline\hline
  Code&$H^{\rm{BDMA}}$ &$\overline{H}_{\rm{SFL}}^{\rm{BDMA}}$ & $H^{\rm{CDMA}}$ &$\overline{H}_{\rm{SFL}}^{\rm{CDMA}}$&$H^{\rm{FDMA}}$ &$\overline{H}_{\rm{SFL}}^{\rm{FDMA}}$&$H^{\rm{DFA}}$ &$\overline{H}_{\rm{SFL}}^{\rm{DFA}}$ &&Code&$H^{\rm{BDMA}}$ &$\overline{H}_{\rm{SFL}}^{\rm{BDMA}}$ & $H^{\rm{CDMA}}$ &$\overline{H}_{\rm{SFL}}^{\rm{CDMA}}$&$H^{\rm{FDMA}}$ &$\overline{H}_{\rm{SFL}}^{\rm{FDMA}}$&$H^{\rm{DFA}}$ &$\overline{H}_{\rm{SFL}}^{\rm{DFA}}$ \\
   \hline
 0001&0.920 &0.501 &0.738 &0.498 &0.916 &0.501 &0.732 &0.500 &&0050 &0.807 &0.502 &0.586 &0.499 &0.802 &0.501 &0.606 &0.501\\
 0002&0.926 &0.500 &0.751 &0.498 &0.925 &0.501 &0.755 &0.501 &&0051 &0.826 &0.503 &0.676 &0.497 &0.816 &0.502 &0.682 &0.501\\
 0003&0.882 &0.501 &0.770 &0.500 &0.880 &0.499 &0.766 &0.501 &&0052 &0.936 &0.501 &0.811 &0.499 &0.936 &0.501 &0.812 &0.501\\
 0004&0.840 &0.499 &0.661 &0.497 &0.828 &0.500 &0.664 &0.499 &&0053 &0.792 &0.499 &0.650 &0.497 &0.783 &0.502 &0.649 &0.500\\
 0005&0.824 &0.501 &0.642 &0.497 &0.805 &0.500 &0.657 &0.500 &&0054 &0.775 &0.501 &0.619 &0.497 &0.767 &0.501 &0.625 &0.498\\
 0006&0.917 &0.499 &0.761 &0.498 &0.915 &0.501 &0.765 &0.500 &&0055 &0.778 &0.502 &0.611 &0.498 &0.771 &0.502 &0.617 &0.500\\
 0007&0.889 &0.501 &0.730 &0.497 &0.885 &0.503 &0.733 &0.500 &&0056 &0.945 &0.500 &0.794 &0.500 &0.944 &0.502 &0.770 &0.501\\
 0008&0.924 &0.503 &0.756 &0.498 &0.923 &0.501 &0.756 &0.502 &&0057 &0.936 &0.502 &0.836 &0.498 &0.935 &0.503 &0.821 &0.500\\
 0009&0.868 &0.502 &0.703 &0.498 &0.855 &0.502 &0.703 &0.501 &&0058 &0.948 &0.500 &0.752 &0.498 &0.948 &0.498 &0.739 &0.501\\
 0010&0.790 &0.499 &0.628 &0.497 &0.785 &0.499 &0.631 &0.500 &&0059 &0.921 &0.500 &0.759 &0.497 &0.920 &0.502 &0.771 &0.500\\
 0011&0.779 &0.502 &0.614 &0.499 &0.774 &0.502 &0.621 &0.500 &&0060 &0.722 &0.502 &0.639 &0.497 &0.726 &0.502 &0.630 &0.500\\
 0012&0.911 &0.502 &0.807 &0.497 &0.913 &0.498 &0.796 &0.501 &&0061 &0.934 &0.502 &0.736 &0.497 &0.930 &0.500 &0.756 &0.501\\
 0013&0.632 &0.499 &0.554 &0.497 &0.633 &0.501 &0.559 &0.500 &&0062 &0.892 &0.502 &0.733 &0.498 &0.892 &0.502 &0.737 &0.500\\
 0014&0.799 &0.503 &0.629 &0.498 &0.794 &0.502 &0.630 &0.500 &&0063 &0.810 &0.501 &0.623 &0.497 &0.803 &0.502 &0.628 &0.500\\
 0015&0.797 &0.503 &0.648 &0.498 &0.791 &0.499 &0.640 &0.500 &&0064 &0.795 &0.501 &0.596 &0.497 &0.790 &0.502 &0.603 &0.501\\
 0016&0.791 &0.501 &0.633 &0.499 &0.782 &0.500 &0.636 &0.499 &&0065 &0.777 &0.502 &0.604 &0.499 &0.771 &0.503 &0.607 &0.500\\
 0017&0.794 &0.503 &0.618 &0.497 &0.790 &0.497 &0.615 &0.500 &&0066 &0.641 &0.503 &0.560 &0.498 &0.630 &0.500 &0.559 &0.501\\
 0018&0.780 &0.502 &0.639 &0.497 &0.774 &0.502 &0.640 &0.500 &&0067 &0.746 &0.502 &0.586 &0.497 &0.735 &0.502 &0.596 &0.500\\
 0019&0.917 &0.500 &0.729 &0.497 &0.913 &0.501 &0.727 &0.499 &&0068 &0.874 &0.501 &0.730 &0.497 &0.872 &0.501 &0.731 &0.499\\
 0020&0.901 &0.501 &0.731 &0.497 &0.896 &0.501 &0.746 &0.500 &&0069 &0.753 &0.502 &0.696 &0.499 &0.755 &0.501 &0.698 &0.500\\
 0021&0.893 &0.501 &0.756 &0.499 &0.891 &0.503 &0.755 &0.500 &&0070 &0.886 &0.500 &0.724 &0.498 &0.885 &0.502 &0.722 &0.501\\
 0022&0.926 &0.502 &0.779 &0.498 &0.927 &0.502 &0.772 &0.500 &&0071 &0.886 &0.500 &0.745 &0.499 &0.885 &0.501 &0.734 &0.500\\
 0023&0.905 &0.499 &0.725 &0.497 &0.905 &0.501 &0.727 &0.500 &&0072 &0.878 &0.501 &0.718 &0.498 &0.877 &0.502 &0.709 &0.499\\
 0024&0.798 &0.502 &0.591 &0.498 &0.794 &0.501 &0.599 &0.500 &&0073 &0.804 &0.500 &0.615 &0.497 &0.798 &0.501 &0.607 &0.501\\
 0025&0.737 &0.502 &0.552 &0.498 &0.731 &0.502 &0.558 &0.500 &&0074 &0.971 &0.503 &0.791 &0.497 &0.971 &0.501 &0.793 &0.500\\
 0026&0.810 &0.502 &0.604 &0.498 &0.803 &0.499 &0.617 &0.501 &&0075 &0.808 &0.502 &0.583 &0.497 &0.805 &0.502 &0.587 &0.499\\
 0027&0.704 &0.500 &0.540 &0.498 &0.695 &0.500 &0.546 &0.501 &&0076 &0.883 &0.502 &0.725 &0.499 &0.882 &0.500 &0.712 &0.499\\
 0028&0.751 &0.501 &0.569 &0.497 &0.744 &0.501 &0.581 &0.500 &&0077 &0.808 &0.499 &0.641 &0.497 &0.802 &0.502 &0.648 &0.501\\
 0029&0.800 &0.503 &0.617 &0.497 &0.794 &0.502 &0.625 &0.501 &&0078 &0.827 &0.502 &0.647 &0.498 &0.819 &0.501 &0.656 &0.501\\
 0030&0.769 &0.501 &0.591 &0.498 &0.765 &0.503 &0.593 &0.501 &&0079 &0.833 &0.500 &0.635 &0.498 &0.829 &0.501 &0.633 &0.501\\
 0031&0.971 &0.501 &0.838 &0.499 &0.970 &0.500 &0.838 &0.500 &&0080 &0.825 &0.499 &0.632 &0.498 &0.819 &0.500 &0.629 &0.499\\
 0032&0.970 &0.500 &0.860 &0.498 &0.968 &0.499 &0.857 &0.501 &&0081 &0.924 &0.501 &0.817 &0.499 &0.924 &0.503 &0.812 &0.500\\
 0033&0.974 &0.502 &0.852 &0.498 &0.973 &0.501 &0.848 &0.500 &&0082 &0.900 &0.502 &0.726 &0.497 &0.899 &0.499 &0.715 &0.500\\
 0034&0.925 &0.501 &0.843 &0.498 &0.925 &0.503 &0.850 &0.500 &&0083 &0.902 &0.502 &0.721 &0.497 &0.902 &0.501 &0.712 &0.500\\
 0035&0.938 &0.501 &0.796 &0.500 &0.938 &0.502 &0.792 &0.500 &&0084 &0.824 &0.500 &0.649 &0.497 &0.816 &0.499 &0.649 &0.500\\
 0036&0.904 &0.503 &0.795 &0.499 &0.901 &0.501 &0.786 &0.501 &&0085 &0.824 &0.500 &0.602 &0.497 &0.819 &0.501 &0.604 &0.501\\
 0037&0.922 &0.500 &0.779 &0.497 &0.922 &0.499 &0.785 &0.501 &&0086 &0.859 &0.502 &0.594 &0.499 &0.856 &0.500 &0.590 &0.500\\
 0038&0.879 &0.503 &0.699 &0.499 &0.878 &0.502 &0.699 &0.501 &&0087 &0.829 &0.501 &0.606 &0.498 &0.825 &0.502 &0.610 &0.500\\
 0039&0.823 &0.503 &0.628 &0.497 &0.817 &0.503 &0.635 &0.501 &&0088 &0.834 &0.501 &0.620 &0.499 &0.829 &0.500 &0.605 &0.500\\
 0040&0.821 &0.502 &0.643 &0.497 &0.814 &0.502 &0.642 &0.501 &&0089 &0.720 &0.500 &0.652 &0.498 &0.721 &0.500 &0.649 &0.500\\
 0041&0.894 &0.502 &0.717 &0.498 &0.893 &0.502 &0.704 &0.501 &&0090 &0.860 &0.503 &0.687 &0.499 &0.852 &0.501 &0.678 &0.501\\
 0042&0.816 &0.502 &0.629 &0.497 &0.810 &0.499 &0.633 &0.501 &&0091 &0.808 &0.501 &0.604 &0.497 &0.802 &0.500 &0.618 &0.500\\
 0043&0.738 &0.501 &0.589 &0.497 &0.730 &0.501 &0.590 &0.502 &&0092 &0.740 &0.500 &0.548 &0.498 &0.732 &0.503 &0.558 &0.502\\
 0044&0.867 &0.501 &0.658 &0.500 &0.862 &0.500 &0.667 &0.501 &&0093 &0.739 &0.501 &0.559 &0.498 &0.733 &0.498 &0.564 &0.500\\
 0045&0.779 &0.497 &0.583 &0.497 &0.772 &0.500 &0.588 &0.501 &&0094 &0.758 &0.503 &0.553 &0.499 &0.751 &0.503 &0.562 &0.501\\
 0046&0.667 &0.500 &0.529 &0.498 &0.661 &0.502 &0.535 &0.500 &&0095 &0.765 &0.501 &0.583 &0.498 &0.762 &0.502 &0.585 &0.501\\
 0047&0.614 &0.501 &0.543 &0.498 &0.612 &0.502 &0.541 &0.499 &&0096 &0.863 &0.502 &0.679 &0.498 &0.861 &0.502 &0.669 &0.501\\
 0048&0.741 &0.501 &0.565 &0.497 &0.731 &0.500 &0.567 &0.501 &&0097 &0.862 &0.502 &0.650 &0.497 &0.857 &0.501 &0.655 &0.500\\
 0049&0.701 &0.502 &0.579 &0.498 &0.689 &0.500 &0.579 &0.501 &&0098 &0.844 &0.500 &0.688 &0.497 &0.841 &0.500 &0.678 &0.500\\
   \hline\hline
   \end{tabular}}
\end{table*}

In Table~\ref{TB:DFA:DMA:H}, one can find that the exponent $H^{\rm{BDMA}}$ mainly concentrates in the range $[0.700,~0.950]$ with the mean value $\overline{H}^{\rm{BDMA}}=0.835\pm0.079$, the exponent $H^{\rm{CDMA}}$ mainly fluctuates in the range $[0.550,~0.850]$ with the mean value $\overline{H}^{\rm{CDMA}}=0.670\pm0.085$, the exponent $H^{\rm{FDMA}}$ mainly varies in the range $[0.700,~0.950]$ with the mean value $\overline{H}^{\rm{FDMA}}=0.830\pm0.081$ and the exponent $H^{\rm{DFA}}$ mainly fluctuates in the range $[0.550,~0.850]$ with the mean value $\overline{H}^{\rm{DFA}}=0.672\pm0.082$, respectively.
%In order to compare the results obtained from the four methods, we investigate the relation among the four scaling exponents $H^{\rm{BDMA}}$, $H^{\rm{CDMA}}$, $H^{\rm{FDMA}}$ and $H^{\rm{DFA}}$ for all the 98 duration series $\tau$. For instance, a simple linear regression shows that
%\begin{equation}
%  H^{\rm{DFA}}= a_0 + a_1 H^{\rm{BDMA}},
%\end{equation}
%where $a_0=-0.100$ and $a_1=0.925$ and the adjusted $R$-square is 0.81. The estimated values of the two coefficients $a_0$ and $a_1$ are significantly different from 0 with the $p$-values less than 0.1\%. Then, in a similar way, the significant positive correlations are unveiled among other scaling exponents.
Since all the scaling exponents ($H^{\rm{BDMA}}$, $H^{\rm{CDMA}}$, $H^{\rm{FDMA}}$ and $H^{\rm{DFA}}$) are significantly greater than 0.5 for all 98 inter-donation duration series, we conclude that the inter-donation durations process long memory. This finding provides further evidence that the donation process is a non-Poisson process.
We also found that
\begin{equation}
  H^{\rm{BDMA}} \approx H^{\rm{FDMA}} > H^{\rm{CDMA}}\approx H^{\rm{DFA}}.
\end{equation}
As shown in Fig.~\ref{Fig:XCB:IDD:Fs}, the BDMA and FDMA methods give excellent power-law scalings, whereas the CDMA and DFA methods give evidently worse results. Therefore, we argue that the BDMA and FDMA methods produce more convincing estimates.

Besides, the probability distribution of inter-donation durations $\tau$ may affect its memory effect. In order to uncover the distribution impact, we first shuffle the inter-donation duration series of each virtual world for 100 times, then obtain the shuffled scaling exponents $H_{\rm{SFL}}^{\rm{BDMA}}$, $H_{\rm{SFL}}^{\rm{CDMA}}$, $H_{\rm{SFL}}^{\rm{FDMA}}$ and $H_{\rm{SFL}}^{\rm{DFA}}$ based on the BDMA, CDMA, FDMA and DFA methods, respectively \cite{Gu-Xiong-Zhang-Chen-Zhang-Zhou-2016-CSF}. The average values of 100 shuffled durations series are also illustrated in Table~\ref{TB:DFA:DMA:H}. It is clear that the values of $\overline{H}_{\rm{SFL}}^{\rm{BDMA}}$, $\overline{H}_{\rm{SFL}}^{\rm{CDMA}}$, $\overline{H}_{\rm{SFL}}^{\rm{FDMA}}$ and $\overline{H}_{\rm{SFL}}^{\rm{DFA}}$ for all 98 inter-donation durations series extremely approach to 0.5 which are obviously smaller than the original ones. So we make a conclusion that the probability distribution of durations $\tau$ does not affect the memory effect, and confirm that durations $\tau$ truly process significant long memory for all the 98 virtual worlds.

\section{Multifractal nature}
\label{S1:Multifractal}

We now turn to investigate the possible presence of nonlinear correlations in the inter-donation duration series through adopting the backward MFDMA method which presents a better performance \cite{Gu-Zhou-2010-PRE}. The backward MFDMA method is an extension of the BDMA approach by generalizing the overall fluctuation function in Eq.~(\ref{Eq:F2s}) to the $q$th-order detrended fluctuation function as follow:
\begin{equation}
  F_q(s) = \left\{\frac{1}{N_s}\sum_{v=1}^{N_s} {F_v^q(s)}\right\}^{\frac{1}{q}},
  \label{Eq:Fqs}
\end{equation}
where $q$ can take any real value except for $q=0$. When $q=0$,
we have
\begin{equation}
  \ln[F_0(s)] = \frac{1}{N_s}\sum_{v=1}^{N_s}{\ln[F_v(s)]},
  \label{Eq:Fq0}
\end{equation}
according to L'H\^{o}spital's rule. Varying the values of box size $s$, we can determine the power-law relation between the function $F_q(s)$ and the size scale $s$,
\begin{equation}
  F_q(s) \sim {s}^{h(q)}.
  \label{Eq:hq}
\end{equation}
where $h(q)$ is the  backward MFDMA scaling exponent. When $q=2$, $h(2)$ is exactly the BDMA scaling exponent $H^{\rm{BDMA}}$.

In the standard multifractal formalism, the multifractal scaling exponent $\tau(q)$ can be used to characterize the multifractal nature, which reads
\begin{equation}
\tau(q)=qh(q)-D_f,
\label{Eq:tau:hq}
\end{equation}
where $D_f$ is the fractal dimension of the geometric support of the multifractal measure \cite{Kantelhardt-Zschiegner-KoscielnyBunde-Havlin-Bunde-Stanley-2002-PA}. For inter-donation duration series, we have $D_f=1$. Moreover, it easily obtain the local singularity exponent $\alpha(q)$ and the multifractal spectrum $f(\alpha)$ via the Legendre transform \cite{Halsey-Jensen-Kadanoff-Procaccia-Shraiman-1986-PRA}
\begin{equation}
    \left\{
    \begin{array}{ll}
        \alpha(q)={\rm{d}}\tau(q)/{\rm{d}}q\\
        f(q)=q{\alpha}-{\tau}(q)
    \end{array}
    \right..
\label{Eq:f:alpha:tau}
\end{equation}
Besides, the multifractal spectrum $f(\alpha)$ is directly related to the generalized dimension $D_q$ \cite{Hentschel-Procaccia-1983-PD,Grassberger-1985-PLA,Meneveau-Sreenivasan-1987-PRL}.

\begin{table}[htp]
\setlength\tabcolsep{2.2pt}
\centering
  \caption{The width of the multifractal spectra $\Delta\alpha$ of inter-donation duration series for 98 virtual worlds based on the backward MFDMA method. $\Delta\alpha_{\rm{SFL}}$ is the average spectrum width of 100 shuffled inter-donation duration series and $\gamma = \Delta\alpha - \Delta\alpha_{\rm{SFL}}$. $\Delta D_q$ is the range of generalized dimensions $D_q$.}
  \label{TB:MFBDMA:Dalpha}
   \begin{tabular}{*{11}{c}}
   \hline\hline
  Code&$\Delta\alpha$ &$\Delta\alpha_{\rm{SFL}}$ &$\gamma$ &$\Delta D_q$ && Code&$\Delta\alpha$ &$\Delta\alpha_{\rm{SFL}}$&$\gamma$ &$\Delta D_q$\\
   \hline
 0001&0.47 &0.35 &0.12 &0.38 && 0050&0.51 &0.34 &0.17 &0.41\\
 0002&0.45 &0.33 &0.12 &0.36 && 0051&0.49 &0.37 &0.12 &0.41\\
 0003&0.42 &0.33 &0.09 &0.34 && 0052&0.53 &0.36 &0.17 &0.43\\
 0004&0.48 &0.35 &0.13 &0.39 && 0053&0.48 &0.35 &0.13 &0.39\\
 0005&0.44 &0.34 &0.10 &0.36 && 0054&0.43 &0.33 &0.10 &0.35\\
 0006&0.46 &0.35 &0.11 &0.38 && 0055&0.44 &0.35 &0.09 &0.36\\
 0007&0.53 &0.36 &0.17 &0.43 && 0056&0.50 &0.33 &0.17 &0.41\\
 0008&0.42 &0.32 &0.10 &0.34 && 0057&0.46 &0.35 &0.11 &0.38\\
 0009&0.49 &0.35 &0.14 &0.39 && 0058&0.51 &0.34 &0.17 &0.41\\
 0010&0.49 &0.33 &0.16 &0.39 && 0059&0.53 &0.34 &0.19 &0.43\\
 0011&0.51 &0.33 &0.18 &0.42 && 0060&0.61 &0.30 &0.31 &0.52\\
 0012&0.44 &0.32 &0.12 &0.36 && 0061&0.49 &0.36 &0.13 &0.39\\
 0013&0.66 &0.32 &0.34 &0.56 && 0062&0.57 &0.33 &0.24 &0.45\\
 0014&0.48 &0.34 &0.14 &0.39 && 0063&0.45 &0.35 &0.10 &0.37\\
 0015&0.50 &0.34 &0.16 &0.41 && 0064&0.47 &0.33 &0.14 &0.39\\
 0016&0.46 &0.33 &0.13 &0.38 && 0065&0.43 &0.33 &0.10 &0.35\\
 0017&0.50 &0.35 &0.15 &0.40 && 0066&0.48 &0.32 &0.16 &0.39\\
 0018&0.53 &0.36 &0.17 &0.43 && 0067&0.56 &0.33 &0.23 &0.46\\
 0019&0.52 &0.35 &0.17 &0.43 && 0068&0.48 &0.33 &0.15 &0.38\\
 0020&0.47 &0.35 &0.12 &0.39 && 0069&0.64 &0.28 &0.36 &0.54\\
 0021&0.53 &0.32 &0.21 &0.42 && 0070&0.44 &0.33 &0.11 &0.36\\
 0022&0.45 &0.32 &0.13 &0.37 && 0071&0.53 &0.34 &0.19 &0.44\\
 0023&0.61 &0.33 &0.28 &0.49 && 0072&0.45 &0.34 &0.11 &0.36\\
 0024&0.55 &0.33 &0.22 &0.44 && 0073&0.44 &0.33 &0.11 &0.36\\
 0025&0.57 &0.33 &0.24 &0.46 && 0074&0.49 &0.35 &0.14 &0.40\\
 0026&0.50 &0.33 &0.17 &0.40 && 0075&0.49 &0.35 &0.14 &0.39\\
 0027&0.58 &0.32 &0.26 &0.47 && 0076&0.43 &0.34 &0.09 &0.35\\
 0028&0.52 &0.34 &0.18 &0.44 && 0077&0.48 &0.35 &0.13 &0.39\\
 0029&0.54 &0.35 &0.19 &0.45 && 0078&0.47 &0.33 &0.14 &0.38\\
 0030&0.48 &0.33 &0.15 &0.39 && 0079&0.44 &0.32 &0.12 &0.35\\
 0031&0.42 &0.32 &0.10 &0.34 && 0080&0.48 &0.33 &0.15 &0.39\\
 0032&0.41 &0.31 &0.10 &0.33 && 0081&0.44 &0.33 &0.11 &0.36\\
 0033&0.42 &0.31 &0.11 &0.34 && 0082&0.45 &0.33 &0.12 &0.37\\
 0034&0.45 &0.33 &0.12 &0.37 && 0083&0.48 &0.33 &0.15 &0.39\\
 0035&0.55 &0.35 &0.20 &0.45 && 0084&0.48 &0.35 &0.13 &0.38\\
 0036&0.51 &0.35 &0.16 &0.42 && 0085&0.45 &0.33 &0.12 &0.36\\
 0037&0.53 &0.35 &0.18 &0.42 && 0086&0.43 &0.33 &0.10 &0.35\\
 0038&0.43 &0.32 &0.11 &0.35 && 0087&0.47 &0.34 &0.13 &0.39\\
 0039&0.46 &0.34 &0.12 &0.38 && 0088&0.46 &0.34 &0.12 &0.38\\
 0040&0.46 &0.33 &0.13 &0.38 && 0089&0.66 &0.30 &0.36 &0.55\\
 0041&0.44 &0.32 &0.12 &0.36 && 0090&0.54 &0.36 &0.18 &0.45\\
 0042&0.47 &0.33 &0.14 &0.39 && 0091&0.50 &0.35 &0.15 &0.41\\
 0043&0.61 &0.34 &0.27 &0.50 && 0092&0.55 &0.34 &0.21 &0.45\\
 0044&0.50 &0.36 &0.14 &0.42 && 0093&0.52 &0.34 &0.18 &0.42\\
 0045&0.51 &0.33 &0.18 &0.40 && 0094&0.54 &0.32 &0.22 &0.43\\
 0046&0.53 &0.31 &0.22 &0.44 && 0095&0.52 &0.34 &0.18 &0.44\\
 0047&0.62 &0.31 &0.31 &0.52 && 0096&0.51 &0.33 &0.18 &0.42\\
 0048&0.57 &0.33 &0.24 &0.46 && 0097&0.51 &0.35 &0.16 &0.41\\
 0049&0.61 &0.32 &0.29 &0.51 && 0098&0.48 &0.32 &0.16 &0.39\\
   \hline\hline
   \end{tabular}
\end{table}

\begin{figure}[!htb]
\centering
\includegraphics[width=0.5\linewidth]{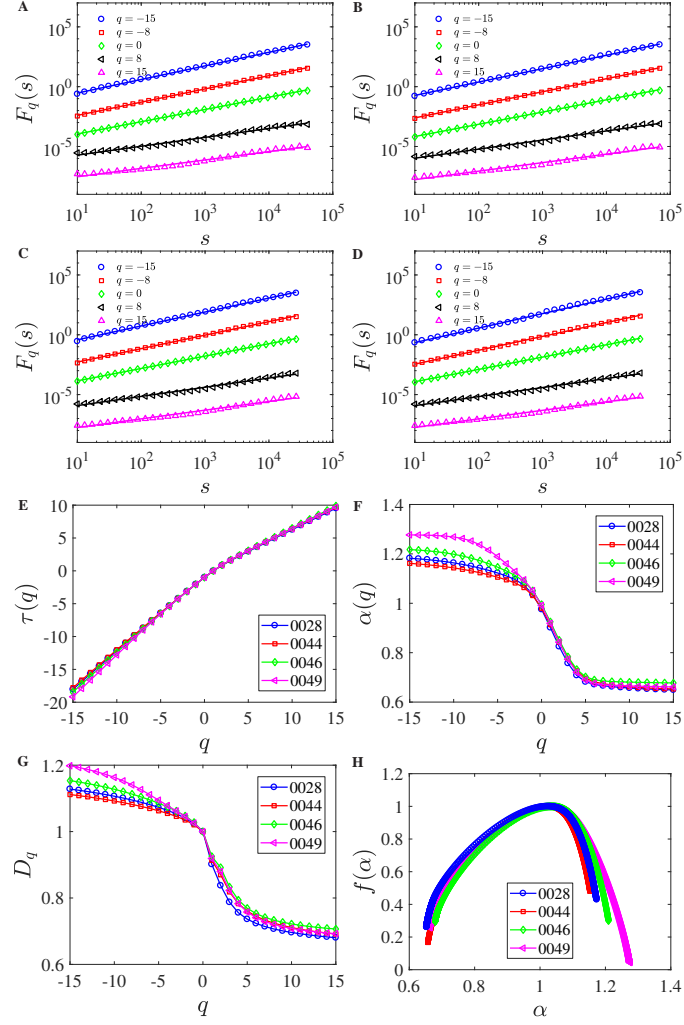}
\caption{Multifractal analysis of the inter-donation durations for four typical virtual worlds. (A-D) Power-law dependence of the fluctuation functions $F_{q}$(s) corresponding to different virtual worlds 0028, 0044, 0046 and 0049. The curves have been shifted vertically for clarity. (E) Multifractal mass exponents $\tau(q)$. (F) Singularity strength functions $\alpha(q)$. (G) Generalized dimensions $D_q$. (H) Multifractal singularity spectra $f(\alpha)$.}
\label{Fig:XCB:IDD:MF}
\end{figure}

Plots (A-D) of Fig.~\ref{Fig:XCB:IDD:MF} depict excellent power-law dependence of the fluctuation functions $F_{q}$(s) with respect to the scale $s$ for four inter-donation duration series. Fig.~\ref{Fig:XCB:IDD:MF} (E) illustrates the nonlinearity of the scaling exponents $\tau(q)$, which suggests the multifractal nature of inter-donation durations. Correspondingly, the singularity strength functions $\alpha(q)$ and the generalized dimensions $D_q$ are presented in Fig.~\ref{Fig:XCB:IDD:MF} (F) and Fig.~\ref{Fig:XCB:IDD:MF} (G), respectively. Additionally, Fig.~\ref{Fig:XCB:IDD:MF} (H) shows the broad multifractal spectra $f(\alpha)$ as a function of the singularity strength $\alpha$ for the same inter-donation duration series in Fig.~\ref{Fig:XCB:IDD:MF} (E). It is well-documented that $\Delta\alpha=\alpha_{\rm{max}}-\alpha_{\rm{min}}$ is an important parameter qualifying the width of multifractal spectrum. The larger the $\Delta\alpha$ value, the stronger the multifractality. As an alternative, $\Delta D_q$ is the range of generalized dimensions $D_q$, which can also quantify multifractality of the inter-donation duration series. The values of $\Delta\alpha$ and $\Delta D_q$ for all the investigated duration series are listed in Table~\ref{TB:MFBDMA:Dalpha}. Table~\ref{TB:MFBDMA:Dalpha} reports that the width of singularity spectrum $\Delta\alpha$ fluctuates in the range $[0.41,0.66]$, which confirms that the inter-donation duration series process multifractality.

\begin{figure}[!htb]
\centering
\includegraphics[width=0.5\linewidth]{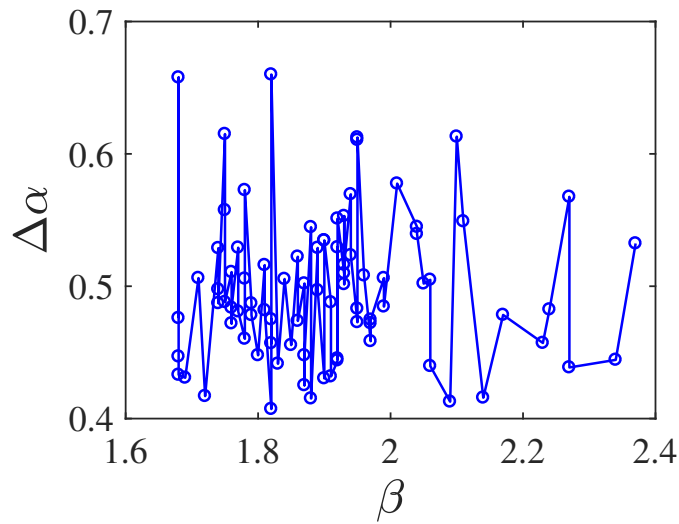}
\caption{Dependence of the singularity width $\Delta \alpha$ as a function of the exponent $\beta$ for the 84 inter-donation duration series with a power-law tail.}
\label{Fig:XCB:IDD:Beta:DeltaAlpha}
\end{figure}

Following the same test procedure in Sec.~\ref{S1:LRC}, we study the impact of probability distribution of inter-donation durations for multifractality \cite{Gu-Xiong-Zhang-Chen-Zhang-Zhou-2016-CSF}. Likewise, we first shuffle the inter-donation duration series for 100 times and then calculate the average value of width of the multifractal spectrum $\Delta\alpha_{\rm{SFL}}$ from 100 shuffled series based on the backward MFDMA method. Since the $\Delta\alpha_{\rm{SFL}}$ reported in Table~\ref{TB:MFBDMA:Dalpha} are obviously larger than zero, we conclude that the distribution of inter-donation durations reliably generates multifractal.
In addition, Fig.~\ref{Fig:XCB:IDD:Beta:DeltaAlpha} illustrates the dependence of the singularity width $\Delta \alpha$ as a function of the exponent $\beta$ for the 84 inter-donation duration series with a power-law tail. More specially, the singularity width $\Delta \alpha$ presents an irregular evolution with the varying exponent $\beta$, which may be attributed to the relatively narrow range of the $\beta$ values.
Correspondingly, we define the residual of spectrum width $\gamma$ through removing the shuffled width $\Delta\alpha_{\rm{SFL}}$ from the original one $\Delta\alpha$, i.e., $\gamma=\Delta\alpha-\Delta\alpha_{\rm{SFL}}$, which provides further evidence that inter-donation durations process multifractal nature (Table~\ref{TB:MFBDMA:Dalpha}).

\section{Conclusion}
\label{S1:Conclusion}

Understanding the regular patterns in individual human interactions is essential in managing information spreading and in tracking social contagion. In this paper, we have investigated the scaling and memory of user donation behaviors based on a popular MMORPG. We presented the characteristic decay of the number of daily user donations for 98 independent virtual worlds inhabited in different servers.
Furthermore, based on the MLE method and the $KS$ statistic \cite{Gu-Xiong-Zhang-Chen-Zhang-Zhou-2016-CSF,Clauset-Shalizi-Newman-2009-SIAMR}, we identified that there are 84 inter-donation duration series following power-law distributions in the tails with an average scaling exponent approaching 1.91.

We then conducted a study on the memory effect of inter-donation duration series by adopting DMA and DFA methods. The empirical investigations suggest that the inter-donation duration series of each virtual world processes strong long-range correlations with the scaling exponent significantly larger than 0.5, and the probability distribution of inter-donation durations has no impact on its memory effect. Finally, we explore the multifractal properties of inter-donation duration series via applying the backward MFDMA algorithm. We found that all the 98 inter-donation duration series exhibit evident multifractal nature and the distribution of inter-donation durations reveals some effect on the multifractality. Summarily, these findings provide essential evidences that the donation process of individuals in virtual worlds cannot be described by the Poisson process.

Our study conducts a systemic investigation of the statistical properties of user donation behaviors (a proxy of direct and indirect reciprocity \cite{Nowak-2005-Science}) in virtual worlds. To the best of our knowledge, similar analysis on donation behavior with such a large database has not been carried out. These empirical findings not only deepen our understanding of the cooperation process, but also can provide stylized facts for the calibration of agent-based models in the field of computational social sciences. It is also possible to construct dynamic models based on Hawkes Processes \cite{Hawkes-1971-Bm,Filimonov-Sornette-2012-PRE} or Markov chains \cite{Jiang-Xie-Li-Zhou-Sornette-2016-JSM} by incorporating the non-Poisson behaviors. In addition, it is worth comparing the scaling and memory of charitable donation behaviors between real and virtual worlds.

\begin{acknowledgments}
 This work was supported by the National Natural Science Foundation of China (11605062, 91746108, 11375064), the China Scholarship Council (201606740024), and the Fundamental Research Funds for the Central Universities (222201818006).
\end{acknowledgments}

%%%\bibliographystyle{spbasic}      % basic style, author-year citations
%%%\bibliographystyle{spmpsci}      % mathematics and physical sciences

%\bibliographystyle{naturemag}
%
%% Create the reference section using BibTeX:
%\bibliography{E:/Auxiliary/Bibliography}

\begin{thebibliography}{10}
\expandafter\ifx\csname url\endcsname\relax
  \def\url#1{\texttt{#1}}\fi
\expandafter\ifx\csname urlprefix\endcsname\relax\def\urlprefix{URL }\fi
\providecommand{\bibinfo}[2]{#2}
\providecommand{\eprint}[2][]{\url{#2}}

\bibitem{OttoniWilhelm-Vesterlund-Xie-2017-AER}
\bibinfo{author}{Ottoni-Wilhelm, M.}, \bibinfo{author}{Vesterlund, L.} \&
  \bibinfo{author}{Xie, H.}
\newblock \bibinfo{title}{{Why do people give? Testing pure and impure
  altruism}}.
\newblock \emph{\bibinfo{journal}{Amer. Econ. Rev.}}
  \textbf{\bibinfo{volume}{107}}, \bibinfo{pages}{3617--3633}
  (\bibinfo{year}{2017}).

\bibitem{Khadjavi-2017-MS}
\bibinfo{author}{Khadjavi, M.}
\newblock \bibinfo{title}{{Indirect reciprocity and charitable giving--evidence
  from a field experiment}}.
\newblock \emph{\bibinfo{journal}{Manag. Sci.}} \textbf{\bibinfo{volume}{63}},
  \bibinfo{pages}{3708--3717} (\bibinfo{year}{2017}).

\bibitem{Nowak-2005-Science}
\bibinfo{author}{Nowak, M.~A.}
\newblock \bibinfo{title}{{Five rules for the evolution of cooperation}}.
\newblock \emph{\bibinfo{journal}{Science}} \textbf{\bibinfo{volume}{314}},
  \bibinfo{pages}{1560--1563} (\bibinfo{year}{2006}).

\bibitem{Nowak-2012-JTB}
\bibinfo{author}{Nowak, M.~A.}
\newblock \bibinfo{title}{{Evolving cooperation}}.
\newblock \emph{\bibinfo{journal}{J. Theor. Biol.}}
  \textbf{\bibinfo{volume}{299}}, \bibinfo{pages}{1--8} (\bibinfo{year}{2012}).

\bibitem{Bainbridge-2007-Science}
\bibinfo{author}{Bainbridge, W.~S.}
\newblock \bibinfo{title}{{The scientific research potential of virtual
  worlds}}.
\newblock \emph{\bibinfo{journal}{Science}} \textbf{\bibinfo{volume}{317}},
  \bibinfo{pages}{472--476} (\bibinfo{year}{2007}).

\bibitem{Grabowski-Kosinski-2008-APPA}
\bibinfo{author}{Grabowski, A.} \& \bibinfo{author}{Kosi{\'n}ski, R.}
\newblock \bibinfo{title}{{The SIRS model of epidemic spreading in virtual
  society}}.
\newblock \emph{\bibinfo{journal}{Acta Phys. Pol. A}}
  \textbf{\bibinfo{volume}{114}}, \bibinfo{pages}{589--596}
  (\bibinfo{year}{2008}).

\bibitem{Jiang-Zhou-Tan-2009-EPL}
\bibinfo{author}{Jiang, Z.-Q.}, \bibinfo{author}{Zhou, W.-X.} \&
  \bibinfo{author}{Tan, Q.-Z.}
\newblock \bibinfo{title}{{Online-offline activities and game-playing behaviors
  of avatars in a massive multiplayer online role-playing game}}.
\newblock \emph{\bibinfo{journal}{EPL (Europhys. Lett.)}}
  \textbf{\bibinfo{volume}{88}}, \bibinfo{pages}{48007} (\bibinfo{year}{2009}).

\bibitem{Jiang-Ren-Gu-Tan-Zhou-2010-PA}
\bibinfo{author}{Jiang, Z.-Q.}, \bibinfo{author}{Ren, F.}, \bibinfo{author}{Gu,
  G.-F.}, \bibinfo{author}{Tan, Q.-Z.} \& \bibinfo{author}{Zhou, W.-X.}
\newblock \bibinfo{title}{{Statistical properties of online avatar numbers in a
  massive multiplayer online role-playing game}}.
\newblock \emph{\bibinfo{journal}{Physica A}} \textbf{\bibinfo{volume}{389}},
  \bibinfo{pages}{807--814} (\bibinfo{year}{2010}).

\bibitem{Xie-Li-Jiang-Tan-Podobnik-Zhou-Stanley-2016-SR}
\bibinfo{author}{Xie, W.-J.} \emph{et~al.}
\newblock \bibinfo{title}{{Skill complementarity enhances heterophily in
  collaboration networks}}.
\newblock \emph{\bibinfo{journal}{Sci. Rep.}} \textbf{\bibinfo{volume}{6}},
  \bibinfo{pages}{18727} (\bibinfo{year}{2016}).

\bibitem{Yang-Xie-Li-Jiang-Zhou-2017-CSF}
\bibinfo{author}{Yang, Y.-H.}, \bibinfo{author}{Xie, W.-J.},
  \bibinfo{author}{Li, M.-X.}, \bibinfo{author}{Jiang, Z.-Q.} \&
  \bibinfo{author}{Zhou, W.-X.}
\newblock \bibinfo{title}{{Statistical properties of user activity fluctuations
  in virtual worlds}}.
\newblock \emph{\bibinfo{journal}{Chaos Solitons Fractals}}
  \textbf{\bibinfo{volume}{105}}, \bibinfo{pages}{271--278}
  (\bibinfo{year}{2017}).

\bibitem{Xie-Yang-Li-Jiang-Zhou-2017-EPJds}
\bibinfo{author}{Xie, W.-J.}, \bibinfo{author}{Yang, Y.-H.},
  \bibinfo{author}{Li, M.-X.}, \bibinfo{author}{Jiang, Z.-Q.} \&
  \bibinfo{author}{Zhou, W.-X.}
\newblock \bibinfo{title}{{Individual position diversity in dependence
  socioeconomic networks increases economic output}}.
\newblock \emph{\bibinfo{journal}{Eur. Phys. J. Data Sci.}}
  \textbf{\bibinfo{volume}{6}}, \bibinfo{pages}{10} (\bibinfo{year}{2017}).

\bibitem{Castronova-2001-WP}
\bibinfo{author}{Castronova, E.}
\newblock \bibinfo{title}{{Virtual worlds: A first-hand account of market and
  society on the cyberian frontier}} (\bibinfo{year}{2001}).
\newblock \bibinfo{note}{Available at SSRN: http://ssrn.com/abstract=294828}.

\bibitem{Xie-Li-Jiang-Zhou-2014-SR}
\bibinfo{author}{Xie, W.-J.}, \bibinfo{author}{Li, M.-X.},
  \bibinfo{author}{Jiang, Z.-Q.} \& \bibinfo{author}{Zhou, W.-X.}
\newblock \bibinfo{title}{{Triadic motifs in the dependence networks of virtual
  societies}}.
\newblock \emph{\bibinfo{journal}{Sci. Rep.}} \textbf{\bibinfo{volume}{4}},
  \bibinfo{pages}{5244} (\bibinfo{year}{2014}).

\bibitem{Szell-Lambiotte-Thurner-2010-PNAS}
\bibinfo{author}{Szell, M.}, \bibinfo{author}{Lambiotte, R.} \&
  \bibinfo{author}{Thurner, S.}
\newblock \bibinfo{title}{{Multirelational organization of large-scale social
  networks in an online world}}.
\newblock \emph{\bibinfo{journal}{Proc. Natl. Acad. Sci. U.S.A.}}
  \textbf{\bibinfo{volume}{107}}, \bibinfo{pages}{13636--13641}
  (\bibinfo{year}{2010}).

\bibitem{Szell-Thurner-2013-SR}
\bibinfo{author}{Szell, M.} \& \bibinfo{author}{Thurner, S.}
\newblock \bibinfo{title}{{How women organize social networks different from
  men}}.
\newblock \emph{\bibinfo{journal}{Sci. Rep.}} \textbf{\bibinfo{volume}{3}},
  \bibinfo{pages}{1214} (\bibinfo{year}{2013}).

\bibitem{Szell-Thurner-2010-SN}
\bibinfo{author}{Szell, M.} \& \bibinfo{author}{Thurner, S.}
\newblock \bibinfo{title}{{Measuring social dynamics in a massive multiplayer
  online game}}.
\newblock \emph{\bibinfo{journal}{Soc. Networks}}
  \textbf{\bibinfo{volume}{32}}, \bibinfo{pages}{313--329}
  (\bibinfo{year}{2010}).

\bibitem{Matsuda-2003-Presence}
\bibinfo{author}{Matsuda, K.}
\newblock \bibinfo{title}{{Can we sell a virtual object in a virtual society?}}
\newblock \emph{\bibinfo{journal}{Presence}} \textbf{\bibinfo{volume}{12}},
  \bibinfo{pages}{581--598} (\bibinfo{year}{2003}).

\bibitem{Castronova-2005-HBR}
\bibinfo{author}{Castronova, E.}
\newblock \bibinfo{title}{{Real products in imaginary worlds}}.
\newblock \emph{\bibinfo{journal}{Harward Buss. Rev.}}
  \textbf{\bibinfo{volume}{83}}, \bibinfo{pages}{20--22}
  (\bibinfo{year}{2005}).

\bibitem{Hemp-2006-HBR}
\bibinfo{author}{Hemp, P.}
\newblock \bibinfo{title}{{Avatar-baed marketing}}.
\newblock \emph{\bibinfo{journal}{Harward Buss. Rev.}}
  \textbf{\bibinfo{volume}{84}}, \bibinfo{pages}{48--57}
  (\bibinfo{year}{2006}).

\bibitem{Papagiannidis-Bourlakis-Li-2008-TFSC}
\bibinfo{author}{Papagiannidis, S.}, \bibinfo{author}{Bourlakis, M.} \&
  \bibinfo{author}{Li, F.}
\newblock \bibinfo{title}{{Making real money in virtual worlds: MMORPGs and
  emerging business opportunities, challenges and ethical implications in
  metaverses}}.
\newblock \emph{\bibinfo{journal}{Tech. Forcast. Soc. Change}}
  \textbf{\bibinfo{volume}{75}}, \bibinfo{pages}{610--622}
  (\bibinfo{year}{2008}).

\bibitem{Zha-Zhou-Zhou-2016-PNAS}
\bibinfo{author}{Zha, Y.-L.}, \bibinfo{author}{Zhou, T.} \&
  \bibinfo{author}{Zhou, C.-S.}
\newblock \bibinfo{title}{{Unfolding large-scale online collaborative human
  dynamics}}.
\newblock \emph{\bibinfo{journal}{Proc. Natl. Acad. Sci. U.S.A.}}
  \textbf{\bibinfo{volume}{113}}, \bibinfo{pages}{14627--14632}
  (\bibinfo{year}{2016}).

\bibitem{Haight-1967}
\bibinfo{author}{Haight, F.~A.}
\newblock \emph{\bibinfo{title}{{Handbook of the Poisson distribution}}}
  (\bibinfo{publisher}{Wiley}, \bibinfo{address}{New York},
  \bibinfo{year}{1967}).

\bibitem{Barabasi-2005-Nature}
\bibinfo{author}{Barab{\'a}si, A.-L.}
\newblock \bibinfo{title}{{The origin of bursts and heavy tails in human
  dynamics}}.
\newblock \emph{\bibinfo{journal}{Nature}} \textbf{\bibinfo{volume}{435}},
  \bibinfo{pages}{207--211} (\bibinfo{year}{2005}).

\bibitem{Malmgren-Stouffer-Motter-Amaral-2008-PNAS}
\bibinfo{author}{Malmgren, R.~D.}, \bibinfo{author}{Stouffer, D.~B.},
  \bibinfo{author}{Motter, A.~E.} \& \bibinfo{author}{Amaral, L. A.~N.}
\newblock \bibinfo{title}{{A Poissonian explanation for heavy tails in e-mail
  communication}}.
\newblock \emph{\bibinfo{journal}{Proc. Natl. Acad. Sci. U.S.A.}}
  \textbf{\bibinfo{volume}{105}}, \bibinfo{pages}{18153--18158}
  (\bibinfo{year}{2008}).

\bibitem{Wu-Zhou-Xiao-Kurths-Schellnhuber-2010-PNAS}
\bibinfo{author}{Wu, Y.}, \bibinfo{author}{Zhou, C.-S.}, \bibinfo{author}{Xiao,
  J.-H.}, \bibinfo{author}{Kurths, J.} \& \bibinfo{author}{Schellnhuber, H.~J.}
\newblock \bibinfo{title}{{Evidence for a bimodal distribution in human
  communication}}.
\newblock \emph{\bibinfo{journal}{Proc. Natl. Acad. Sci. U.S.A.}}
  \textbf{\bibinfo{volume}{107}}, \bibinfo{pages}{18803--18808}
  (\bibinfo{year}{2010}).

\bibitem{Candia-Gonzalez-Wang-Schoenharl-Madey-Barabasi-2008-JPAMT}
\bibinfo{author}{Candia, J.} \emph{et~al.}
\newblock \bibinfo{title}{{Uncovering individual and collective human dynamics
  from mobile phone records}}.
\newblock \emph{\bibinfo{journal}{J. Phys. A}} \textbf{\bibinfo{volume}{41}},
  \bibinfo{pages}{224015} (\bibinfo{year}{2008}).

\bibitem{Jiang-Xie-Li-Podobnik-Zhou-Stanley-2013-PNAS}
\bibinfo{author}{Jiang, Z.-Q.} \emph{et~al.}
\newblock \bibinfo{title}{{Calling patterns in human communication dynamics}}.
\newblock \emph{\bibinfo{journal}{Proc. Natl. Acad. Sci. U.S.A.}}
  \textbf{\bibinfo{volume}{110}}, \bibinfo{pages}{1600--1605}
  (\bibinfo{year}{2013}).

\bibitem{Oliveira-Barabasi-2005-Nature}
\bibinfo{author}{Oliveira, J.~G.} \& \bibinfo{author}{Barab{\'a}si, A.-L.}
\newblock \bibinfo{title}{{Darwin and Einstein correspondence patterns}}.
\newblock \emph{\bibinfo{journal}{Nature}} \textbf{\bibinfo{volume}{437}},
  \bibinfo{pages}{1251} (\bibinfo{year}{2005}).

\bibitem{Malmgren-Stouffer-Campanharo-Amaral-2009-Science}
\bibinfo{author}{Malmgren, R.~D.}, \bibinfo{author}{Stouffer, D.~B.},
  \bibinfo{author}{Campanharo, A. S. L.~O.} \& \bibinfo{author}{Amaral, L.
  A.~N.}
\newblock \bibinfo{title}{{On universality in human correspondence activity}}.
\newblock \emph{\bibinfo{journal}{Science}} \textbf{\bibinfo{volume}{325}},
  \bibinfo{pages}{1696--1700} (\bibinfo{year}{2009}).

\bibitem{Jiang-Chen-Zhou-2008-PA}
\bibinfo{author}{Jiang, Z.-Q.}, \bibinfo{author}{Chen, W.} \&
  \bibinfo{author}{Zhou, W.-X.}
\newblock \bibinfo{title}{{Scaling in the distribution of intertrade durations
  of Chinese stocks}}.
\newblock \emph{\bibinfo{journal}{Physica A}} \textbf{\bibinfo{volume}{387}},
  \bibinfo{pages}{5818--5825} (\bibinfo{year}{2008}).

\bibitem{Gu-Zhou-2009-EPL}
\bibinfo{author}{Gu, G.-F.} \& \bibinfo{author}{Zhou, W.-X.}
\newblock \bibinfo{title}{{Emergence of long memory in stock volatility from a
  modified Mike-Farmer model}}.
\newblock \emph{\bibinfo{journal}{EPL (Europhys. Lett.)}}
  \textbf{\bibinfo{volume}{86}}, \bibinfo{pages}{48002} (\bibinfo{year}{2009}).

\bibitem{Ni-Jiang-Gu-Ren-Chen-Zhou-2010-PA}
\bibinfo{author}{Ni, X.-H.} \emph{et~al.}
\newblock \bibinfo{title}{{Scaling and memory in the non-Poisson process of
  limit order cancelation}}.
\newblock \emph{\bibinfo{journal}{Physica A}} \textbf{\bibinfo{volume}{389}},
  \bibinfo{pages}{2751--2761} (\bibinfo{year}{2010}).

\bibitem{Ruan-Zhou-2011-PA}
\bibinfo{author}{Ruan, Y.-P.} \& \bibinfo{author}{Zhou, W.-X.}
\newblock \bibinfo{title}{{Long-term correlations and multifractal nature in
  the intertrade durations of a liquid Chinese stock and its warrant}}.
\newblock \emph{\bibinfo{journal}{Physica A}} \textbf{\bibinfo{volume}{390}},
  \bibinfo{pages}{1646--1654} (\bibinfo{year}{2011}).

\bibitem{Gu-Xiong-Zhang-Zhang-Zhou-2014-FiP}
\bibinfo{author}{Gu, G.-F.}, \bibinfo{author}{Xiong, X.},
  \bibinfo{author}{Zhang, W.}, \bibinfo{author}{Zhang, Y.-J.} \&
  \bibinfo{author}{Zhou, W.-X.}
\newblock \bibinfo{title}{{Empirical properties of inter-cancellation durations
  in the Chinese stock market}}.
\newblock \emph{\bibinfo{journal}{Front. in Phys.}}
  \textbf{\bibinfo{volume}{2}}, \bibinfo{pages}{16} (\bibinfo{year}{2014}).

\bibitem{Sornette-2004}
\bibinfo{author}{Sornette, D.}
\newblock \emph{\bibinfo{title}{{Critical Phenomena in Natural Sciences}}}
  (\bibinfo{publisher}{Springer}, \bibinfo{address}{Berlin},
  \bibinfo{year}{2004}), \bibinfo{edition}{2} edn.

\bibitem{Gu-Xiong-Zhang-Chen-Zhang-Zhou-2016-CSF}
\bibinfo{author}{Gu, G.-F.} \emph{et~al.}
\newblock \bibinfo{title}{{Stylized facts of price gaps in limit order books}}.
\newblock \emph{\bibinfo{journal}{Chaos Solitons Fractals}}
  \textbf{\bibinfo{volume}{88}}, \bibinfo{pages}{48--58}
  (\bibinfo{year}{2016}).

\bibitem{Peng-Buldyrev-Goldberger-Havlin-Sciortino-Simons-Stanley-1992-Nature}
\bibinfo{author}{Peng, C.-K.} \emph{et~al.}
\newblock \bibinfo{title}{{Long-range correlations in nucleotide sequences}}.
\newblock \emph{\bibinfo{journal}{Nature}} \textbf{\bibinfo{volume}{356}},
  \bibinfo{pages}{168--170} (\bibinfo{year}{1992}).

\bibitem{Bunde-Bunde-Havlin-Roman-Goldreich-Schellnhuber-1998-PRL}
\bibinfo{author}{Koscielny-Bunde, E.} \emph{et~al.}
\newblock \bibinfo{title}{{Indication of a universal persistence law governing
  atmospheric variability}}.
\newblock \emph{\bibinfo{journal}{Phys. Rev. Lett.}}
  \textbf{\bibinfo{volume}{81}}, \bibinfo{pages}{729--732}
  (\bibinfo{year}{1998}).

\bibitem{Rybski-Buldyrev-Havlin-Liljeros-Makse-2009-PNAS}
\bibinfo{author}{Rybski, D.}, \bibinfo{author}{Buldyrev, S.~V.},
  \bibinfo{author}{Havlin, S.}, \bibinfo{author}{Liljeros, F.} \&
  \bibinfo{author}{Makse, H.~A.}
\newblock \bibinfo{title}{{Scaling laws of human interaction activity}}.
\newblock \emph{\bibinfo{journal}{Proc. Natl. Acad. Sci. U.S.A.}}
  \textbf{\bibinfo{volume}{106}}, \bibinfo{pages}{12640--12645}
  (\bibinfo{year}{2009}).

\bibitem{Rozenfeld-Rybski-Andrade-Batty-Stanley-Makse-2008-PNAS}
\bibinfo{author}{Rozenfeld, H.~D.} \emph{et~al.}
\newblock \bibinfo{title}{{Laws of population growth}}.
\newblock \emph{\bibinfo{journal}{Proc. Natl. Acad. Sci. U.S.A.}}
  \textbf{\bibinfo{volume}{105}}, \bibinfo{pages}{18702--18707}
  (\bibinfo{year}{2008}).

\bibitem{Taqqu-Teverovsky-Willinger-1995-Fractals}
\bibinfo{author}{Taqqu, M.~S.}, \bibinfo{author}{Teverovsky, V.} \&
  \bibinfo{author}{Willinger, W.}
\newblock \bibinfo{title}{{Estimators for long-range dependence: An empirical
  study}}.
\newblock \emph{\bibinfo{journal}{Fractals}} \textbf{\bibinfo{volume}{3}},
  \bibinfo{pages}{785--798} (\bibinfo{year}{1995}).

\bibitem{Montanari-Taqqu-Teverovsky-1999-MCM}
\bibinfo{author}{Montanari, A.}, \bibinfo{author}{Taqqu, M.~S.} \&
  \bibinfo{author}{Teverovsky, V.}
\newblock \bibinfo{title}{{Estimating long-range dependence in the presence of
  periodicity: An empirical study}}.
\newblock \emph{\bibinfo{journal}{Math. Comput. Model.}}
  \textbf{\bibinfo{volume}{29}}, \bibinfo{pages}{217--228}
  (\bibinfo{year}{1999}).

\bibitem{Delignieres-Ramdani-Lemoine-Torre-Fortes-Ninot-2006-JMPsy}
\bibinfo{author}{Delignieres, D.} \emph{et~al.}
\newblock \bibinfo{title}{{Fractal analyses for `short' time series: A
  re-assessment of classical methods}}.
\newblock \emph{\bibinfo{journal}{J. Math. Psychol.}}
  \textbf{\bibinfo{volume}{50}}, \bibinfo{pages}{525--544}
  (\bibinfo{year}{2006}).

\bibitem{Kantelhardt-2009-ECSS}
\bibinfo{author}{Kantelhardt, J.~W.}
\newblock \bibinfo{title}{Fractal and multifractal time series}.
\newblock In \bibinfo{editor}{Meyers, R.~A.} (ed.)
  \emph{\bibinfo{booktitle}{Encyclopedia of Complexity and Systems Science}},
  vol. \bibinfo{volume}{LXXX}, \bibinfo{pages}{3754--3778}
  (\bibinfo{publisher}{Springer}, \bibinfo{address}{Berlin},
  \bibinfo{year}{2009}).

\bibitem{Hurst-1951-TASCE}
\bibinfo{author}{Hurst, H.~E.}
\newblock \bibinfo{title}{{Long-term storage capacity of reservoirs}}.
\newblock \emph{\bibinfo{journal}{Trans. Amer. Soc. Civil Eng.}}
  \textbf{\bibinfo{volume}{116}}, \bibinfo{pages}{770--808}
  (\bibinfo{year}{1951}).

\bibitem{Bunde-Eichner-Havlin-Kantelhardt-2004-PA}
\bibinfo{author}{Bunde, A.}, \bibinfo{author}{Eichner, J.~F.},
  \bibinfo{author}{Havlin, S.} \& \bibinfo{author}{Kantelhardt, J.~W.}
\newblock \bibinfo{title}{{Return intervals of rare events in records with
  long-term persistence}}.
\newblock \emph{\bibinfo{journal}{Physica A}} \textbf{\bibinfo{volume}{342}},
  \bibinfo{pages}{308--314} (\bibinfo{year}{2004}).

\bibitem{Kantelhardt-Roman-Greiner-1995-PA}
\bibinfo{author}{Kantelhardt, J.~W.}, \bibinfo{author}{Roman, H.~E.} \&
  \bibinfo{author}{Greiner, M.}
\newblock \bibinfo{title}{{Discrete wavelet approach to multifractality}}.
\newblock \emph{\bibinfo{journal}{Physica A}} \textbf{\bibinfo{volume}{220}},
  \bibinfo{pages}{219--238} (\bibinfo{year}{1995}).

\bibitem{Holschneider-1988-JSP}
\bibinfo{author}{Holschneider, M.}
\newblock \bibinfo{title}{{On the wavelet transformation of fractal objects}}.
\newblock \emph{\bibinfo{journal}{J. Stat. Phys.}}
  \textbf{\bibinfo{volume}{50}}, \bibinfo{pages}{963--993}
  (\bibinfo{year}{1988}).

\bibitem{Muzy-Bacry-Arneodo-1991-PRL}
\bibinfo{author}{Muzy, J.~F.}, \bibinfo{author}{Bacry, E.} \&
  \bibinfo{author}{Arn{\'e}odo, A.}
\newblock \bibinfo{title}{{Wavelets and multifractal formalism for singular
  signals: Application to turbulence data}}.
\newblock \emph{\bibinfo{journal}{Phys. Rev. Lett.}}
  \textbf{\bibinfo{volume}{67}}, \bibinfo{pages}{3515--3518}
  (\bibinfo{year}{1991}).

\bibitem{Bacry-Muzy-Arneodo-1993-JSP}
\bibinfo{author}{Bacry, E.}, \bibinfo{author}{Muzy, J.~F.} \&
  \bibinfo{author}{Arn{\'e}odo, A.}
\newblock \bibinfo{title}{{Singularity spectrum of fractal signals from wavelet
  analysis: Exact results}}.
\newblock \emph{\bibinfo{journal}{J. Stat. Phys.}}
  \textbf{\bibinfo{volume}{70}}, \bibinfo{pages}{635--674}
  (\bibinfo{year}{1993}).

\bibitem{Muzy-Bacry-Arneodo-1993-PRE}
\bibinfo{author}{Muzy, J.~F.}, \bibinfo{author}{Bacry, E.} \&
  \bibinfo{author}{Arn{\'e}odo, A.}
\newblock \bibinfo{title}{{Multifractal formalism for fractal signals: The
  structure-function approach versus the wavelet-transform modulus-maxima
  method}}.
\newblock \emph{\bibinfo{journal}{Phys. Rev. E}} \textbf{\bibinfo{volume}{47}},
  \bibinfo{pages}{875--884} (\bibinfo{year}{1993}).

\bibitem{Muzy-Bacry-Arneodo-1994-IJBC}
\bibinfo{author}{Muzy, J.~F.}, \bibinfo{author}{Bacry, E.} \&
  \bibinfo{author}{Arn{\'e}{o}do, A.}
\newblock \bibinfo{title}{{The multifractal formalism revisited with
  wavelets}}.
\newblock \emph{\bibinfo{journal}{Int. J. Bifurcation Chaos}}
  \textbf{\bibinfo{volume}{4}}, \bibinfo{pages}{245--302}
  (\bibinfo{year}{1994}).

\bibitem{Peng-Buldyrev-Havlin-Simons-Stanley-Goldberger-1994-PRE}
\bibinfo{author}{Peng, C.-K.} \emph{et~al.}
\newblock \bibinfo{title}{{Mosaic organization of DNA nucleotides}}.
\newblock \emph{\bibinfo{journal}{Phys. Rev. E}} \textbf{\bibinfo{volume}{49}},
  \bibinfo{pages}{1685--1689} (\bibinfo{year}{1994}).

\bibitem{Alessio-Carbone-Castelli-Frappietro-2002-EPJB}
\bibinfo{author}{Alessio, E.}, \bibinfo{author}{Carbone, A.},
  \bibinfo{author}{Castelli, G.} \& \bibinfo{author}{Frappietro, V.}
\newblock \bibinfo{title}{{Second-order moving average and scaling of
  stochastic time series}}.
\newblock \emph{\bibinfo{journal}{Eur. Phys. J. B}}
  \textbf{\bibinfo{volume}{27}}, \bibinfo{pages}{197--200}
  (\bibinfo{year}{2002}).

\bibitem{Carbone-Castelli-2003-SPIE}
\bibinfo{author}{Carbone, A.} \& \bibinfo{author}{Castelli, G.}
\newblock \bibinfo{title}{{Scaling properties of long-range correlated noisy
  signals: Appplication to financial markets}}.
\newblock \emph{\bibinfo{journal}{Proc. SPIE}} \textbf{\bibinfo{volume}{5114}},
  \bibinfo{pages}{406--414} (\bibinfo{year}{2003}).

\bibitem{Xu-Ivanov-Hu-Chen-Carbone-Stanley-2005-PRE}
\bibinfo{author}{Xu, L.~M.} \emph{et~al.}
\newblock \bibinfo{title}{{Quantifying signals with power-law correlations: A
  comparative study of detrended fluctuation analysis and detrended moving
  average techniques}}.
\newblock \emph{\bibinfo{journal}{Phys. Rev. E}} \textbf{\bibinfo{volume}{71}},
  \bibinfo{pages}{051101} (\bibinfo{year}{2005}).

\bibitem{Oswiecimka-Kwapien-Drozdz-2006-PRE}
\bibinfo{author}{O{\'s}wi{\c{e}}cimka, P.}, \bibinfo{author}{Kwapie{\'n}, J.}
  \& \bibinfo{author}{Dro$\dot{z}$d$\dot{z}$, S.}
\newblock \bibinfo{title}{{Wavelet versus detrended fluctuation analysis of
  multifractal structures}}.
\newblock \emph{\bibinfo{journal}{Phys. Rev. E}} \textbf{\bibinfo{volume}{74}},
  \bibinfo{pages}{016103} (\bibinfo{year}{2006}).

\bibitem{Bashan-Bartsch-Kantelhardt-Havlin-2008-PA}
\bibinfo{author}{Bashan, A.}, \bibinfo{author}{Bartsch, R.},
  \bibinfo{author}{Kantelhardt, J.~W.} \& \bibinfo{author}{Havlin, S.}
\newblock \bibinfo{title}{{Comparison of detrending methods for fluctuation
  analysis}}.
\newblock \emph{\bibinfo{journal}{Physica A}} \textbf{\bibinfo{volume}{387}},
  \bibinfo{pages}{5080--5090} (\bibinfo{year}{2008}).

\bibitem{Serinaldi-2010-PA}
\bibinfo{author}{Serinaldi, F.}
\newblock \bibinfo{title}{{Use and misuse of some Hurst parameter estimators
  applied to stationary and non-stationary financial time series}}.
\newblock \emph{\bibinfo{journal}{Physica A}} \textbf{\bibinfo{volume}{389}},
  \bibinfo{pages}{2770--2781} (\bibinfo{year}{2010}).

\bibitem{Jiang-Zhou-2011-PRE}
\bibinfo{author}{Jiang, Z.-Q.} \& \bibinfo{author}{Zhou, W.-X.}
\newblock \bibinfo{title}{{Multifractal detrending moving-average
  cross-correlation analysis}}.
\newblock \emph{\bibinfo{journal}{Phys. Rev. E}} \textbf{\bibinfo{volume}{84}},
  \bibinfo{pages}{016106} (\bibinfo{year}{2011}).

\bibitem{Huang-Schmitt-Hermand-Gagne-Lu-Liu-2011-PRE}
\bibinfo{author}{Huang, Y.-X.} \emph{et~al.}
\newblock \bibinfo{title}{{Arbitrary-order Hilbert spectral analysis for time
  series possessing scaling statistics: Comparison study with detrended
  fluctuation analysis and wavelet leaders}}.
\newblock \emph{\bibinfo{journal}{Phys. Rev. E}} \textbf{\bibinfo{volume}{84}},
  \bibinfo{pages}{016208} (\bibinfo{year}{2011}).

\bibitem{Bryce-Sprague-2012-SR}
\bibinfo{author}{Bryce, R.~M.} \& \bibinfo{author}{Sprague, K.~B.}
\newblock \bibinfo{title}{{Revisiting detrended fluctuation analysis}}.
\newblock \emph{\bibinfo{journal}{Sci. Rep.}} \textbf{\bibinfo{volume}{2}},
  \bibinfo{pages}{315} (\bibinfo{year}{2012}).

\bibitem{Shao-Gu-Jiang-Zhou-2015-Fractals}
\bibinfo{author}{Shao, Y.-H.}, \bibinfo{author}{Gu, G.-F.},
  \bibinfo{author}{Jiang, Z.-Q.} \& \bibinfo{author}{Zhou, W.-X.}
\newblock \bibinfo{title}{{Effects of polynomial trends on detrending moving
  average analysis}}.
\newblock \emph{\bibinfo{journal}{Fractals}} \textbf{\bibinfo{volume}{23}},
  \bibinfo{pages}{1550034} (\bibinfo{year}{2015}).

\bibitem{Grech-Mazur-2013-PRE}
\bibinfo{author}{Grech, D.} \& \bibinfo{author}{Mazur, Z.}
\newblock \bibinfo{title}{{Scaling range of power-laws that originate from
  fluctuation analysis}}.
\newblock \emph{\bibinfo{journal}{Phys. Rev. E}} \textbf{\bibinfo{volume}{87}},
  \bibinfo{pages}{052809} (\bibinfo{year}{2013}).

\bibitem{Grech-Mazur-2015-APPA}
\bibinfo{author}{Grech, D.} \& \bibinfo{author}{Mazur, Z.}
\newblock \bibinfo{title}{{Impact of scaling range on the effectiveness of
  detrending methods}}.
\newblock \emph{\bibinfo{journal}{Acta Phys. Pol. A}}
  \textbf{\bibinfo{volume}{127}}, \bibinfo{pages}{A59--A65}
  (\bibinfo{year}{2015}).

\bibitem{Shao-Gu-Jiang-Zhou-Sornette-2012-SR}
\bibinfo{author}{Shao, Y.-H.}, \bibinfo{author}{Gu, G.-F.},
  \bibinfo{author}{Jiang, Z.-Q.}, \bibinfo{author}{Zhou, W.-X.} \&
  \bibinfo{author}{Sornette, D.}
\newblock \bibinfo{title}{{Comparing the performance of FA, DFA and DMA using
  different synthetic long-range correlated time series}}.
\newblock \emph{\bibinfo{journal}{Sci. Rep.}} \textbf{\bibinfo{volume}{2}},
  \bibinfo{pages}{835} (\bibinfo{year}{2012}).

\bibitem{Mandelbrot-1983}
\bibinfo{author}{Mandelbrot, B.~B.}
\newblock \emph{\bibinfo{title}{{The Fractal Geometry of Nature}}}
  (\bibinfo{publisher}{W. H. Freeman}, \bibinfo{address}{New York},
  \bibinfo{year}{1983}).

\bibitem{Jiang-Xie-Zhou-Sornette-2018-XXX}
\bibinfo{author}{Jiang, Z.-Q.}, \bibinfo{author}{Xie, W.-J.},
  \bibinfo{author}{Zhou, W.-X.} \& \bibinfo{author}{Sornette, D.}
\newblock \bibinfo{title}{{Multifractal analysis of financial markets}}.
\newblock \emph{\bibinfo{journal}{arXiv: 1805.04750}}
  \bibinfo{pages}{submitted} (\bibinfo{year}{2018}).

\bibitem{Kolmogorov-1962-JFM}
\bibinfo{author}{Kolmogorov, A.~N.}
\newblock \bibinfo{title}{{A refinement of previous hypotheses concerning the
  local structure of turbulence in a viscous incompressible fluid at high
  Reynolds number}}.
\newblock \emph{\bibinfo{journal}{J. Fluid Mech.}}
  \textbf{\bibinfo{volume}{13}}, \bibinfo{pages}{82--85}
  (\bibinfo{year}{1962}).

\bibitem{VanAtta-Chen-1970-JFM}
\bibinfo{author}{Van~Atta, C.~W.} \& \bibinfo{author}{Chen, W.~Y.}
\newblock \bibinfo{title}{{Structure functions of turbulence in the atmospheric
  boundary layer over the ocean}}.
\newblock \emph{\bibinfo{journal}{J. Fluid Mech.}}
  \textbf{\bibinfo{volume}{44}}, \bibinfo{pages}{145--159}
  (\bibinfo{year}{1970}).

\bibitem{Anselmet-Gagne-Hopfinger-Antonia-1984-JFM}
\bibinfo{author}{Anselmet, F.}, \bibinfo{author}{Gagne, Y.},
  \bibinfo{author}{Hopfinger, E.~J.} \& \bibinfo{author}{Antonia, R.~A.}
\newblock \bibinfo{title}{{High-order velocity structure functions in turbulent
  shear flows}}.
\newblock \emph{\bibinfo{journal}{J. Fluid Mech.}}
  \textbf{\bibinfo{volume}{140}}, \bibinfo{pages}{63--89}
  (\bibinfo{year}{1984}).

\bibitem{Ghashghaie-Breymann-Peinke-Talkner-Dodge-1996-Nature}
\bibinfo{author}{Ghashghaie, S.}, \bibinfo{author}{Breymann, W.},
  \bibinfo{author}{Peinke, J.}, \bibinfo{author}{Talkner, P.} \&
  \bibinfo{author}{Dodge, Y.}
\newblock \bibinfo{title}{{Turbulent cascades in foreign exchange markets}}.
\newblock \emph{\bibinfo{journal}{Nature}} \textbf{\bibinfo{volume}{381}},
  \bibinfo{pages}{767--770} (\bibinfo{year}{1996}).

\bibitem{Grassberger-1983-PLA}
\bibinfo{author}{Grassberger, P.}
\newblock \bibinfo{title}{{Generalized dimensions of strange attractors}}.
\newblock \emph{\bibinfo{journal}{Phys. Lett. A}}
  \textbf{\bibinfo{volume}{97}}, \bibinfo{pages}{227--230}
  (\bibinfo{year}{1983}).

\bibitem{Hentschel-Procaccia-1983-PD}
\bibinfo{author}{Hentschel, H. G.~E.} \& \bibinfo{author}{Procaccia, I.}
\newblock \bibinfo{title}{{The infinite number of generalized dimensions of
  fractals and strange attractors}}.
\newblock \emph{\bibinfo{journal}{Physica D}} \textbf{\bibinfo{volume}{8}},
  \bibinfo{pages}{435--444} (\bibinfo{year}{1983}).

\bibitem{Grassberger-1985-PLA}
\bibinfo{author}{Grassberger, P.}
\newblock \bibinfo{title}{{Generalizations of the Hausdorff dimension of
  fractal measure}}.
\newblock \emph{\bibinfo{journal}{Phys. Lett. A}}
  \textbf{\bibinfo{volume}{107}}, \bibinfo{pages}{101--105}
  (\bibinfo{year}{1985}).

\bibitem{Halsey-Jensen-Kadanoff-Procaccia-Shraiman-1986-PRA}
\bibinfo{author}{Halsey, T.~C.}, \bibinfo{author}{Jensen, M.~H.},
  \bibinfo{author}{Kadanoff, L.~P.}, \bibinfo{author}{Procaccia, I.} \&
  \bibinfo{author}{Shraiman, B.~I.}
\newblock \bibinfo{title}{{Fractal measures and their singularities: The
  characterization of strange sets}}.
\newblock \emph{\bibinfo{journal}{Phys. Rev. A}} \textbf{\bibinfo{volume}{33}},
  \bibinfo{pages}{1141--1151} (\bibinfo{year}{1986}).

\bibitem{Xie-Jiang-Gu-Xiong-Zhou-2015-NJP}
\bibinfo{author}{Xie, W.-J.}, \bibinfo{author}{Jiang, Z.-Q.},
  \bibinfo{author}{Gu, G.-F.}, \bibinfo{author}{Xiong, X.} \&
  \bibinfo{author}{Zhou, W.-X.}
\newblock \bibinfo{title}{{Joint multifractal analysis based on the partition
  function approach: Analytical analysis, numerical simulations and empirical
  application}}.
\newblock \emph{\bibinfo{journal}{New J. Phys.}} \textbf{\bibinfo{volume}{17}},
  \bibinfo{pages}{103020} (\bibinfo{year}{2015}).

\bibitem{Chhabra-Sreenivasan-1992-PRL}
\bibinfo{author}{Chhabra, A.~B.} \& \bibinfo{author}{Sreenivasan, K.~R.}
\newblock \bibinfo{title}{{Scale-invariant multiplier distribution in
  turbulence}}.
\newblock \emph{\bibinfo{journal}{Phys. Rev. Lett.}}
  \textbf{\bibinfo{volume}{68}}, \bibinfo{pages}{2762--2765}
  (\bibinfo{year}{1992}).

\bibitem{Jouault-Lipa-Greiner-1999-PRE}
\bibinfo{author}{Jouault, B.}, \bibinfo{author}{Lipa, P.} \&
  \bibinfo{author}{Greiner, M.}
\newblock \bibinfo{title}{{Multiplier phenomenology in random multiplicative
  cascade processses}}.
\newblock \emph{\bibinfo{journal}{Phys. Rev. E}} \textbf{\bibinfo{volume}{59}},
  \bibinfo{pages}{2451--2454} (\bibinfo{year}{1999}).

\bibitem{Jiang-Zhou-2007-PA}
\bibinfo{author}{Jiang, Z.-Q.} \& \bibinfo{author}{Zhou, W.-X.}
\newblock \bibinfo{title}{{Scale invariant distribution and multifractality of
  volatility multipliers in stock markets}}.
\newblock \emph{\bibinfo{journal}{Physica A}} \textbf{\bibinfo{volume}{381}},
  \bibinfo{pages}{343--350} (\bibinfo{year}{2007}).

\bibitem{Gu-Zhou-2010-PRE}
\bibinfo{author}{Gu, G.-F.} \& \bibinfo{author}{Zhou, W.-X.}
\newblock \bibinfo{title}{{Detrending moving average algorithm for
  multifractals}}.
\newblock \emph{\bibinfo{journal}{Phys. Rev. E}} \textbf{\bibinfo{volume}{82}},
  \bibinfo{pages}{011136} (\bibinfo{year}{2010}).

\bibitem{CastroESilva-Moreira-1997-PA}
\bibinfo{author}{Castro~e Silva, A.} \& \bibinfo{author}{Moreira, J.~G.}
\newblock \bibinfo{title}{{Roughness exponents to calculate multi-affine
  fractal exponents}}.
\newblock \emph{\bibinfo{journal}{Physica A}} \textbf{\bibinfo{volume}{235}},
  \bibinfo{pages}{327--333} (\bibinfo{year}{1997}).

\bibitem{Weber-Talkner-2001-JGR}
\bibinfo{author}{Weber, R.~O.} \& \bibinfo{author}{Talkner, P.}
\newblock \bibinfo{title}{{Spectra and correlations of climate data from days
  to decades}}.
\newblock \emph{\bibinfo{journal}{J. Geophys. Res.}}
  \textbf{\bibinfo{volume}{106}}, \bibinfo{pages}{20131--20144}
  (\bibinfo{year}{2001}).

\bibitem{Kantelhardt-Zschiegner-KoscielnyBunde-Havlin-Bunde-Stanley-2002-PA}
\bibinfo{author}{Kantelhardt, J.~W.} \emph{et~al.}
\newblock \bibinfo{title}{{Multifractal detrended fluctuation analysis of
  nonstationary time series}}.
\newblock \emph{\bibinfo{journal}{Physica A}} \textbf{\bibinfo{volume}{316}},
  \bibinfo{pages}{87--114} (\bibinfo{year}{2002}).

\bibitem{Zhou-2012-CSF}
\bibinfo{author}{Zhou, W.-X.}
\newblock \bibinfo{title}{{Finite-size effect and the components of
  multifractality in financial volatility}}.
\newblock \emph{\bibinfo{journal}{Chaos Solitons Fractals}}
  \textbf{\bibinfo{volume}{45}}, \bibinfo{pages}{147--155}
  (\bibinfo{year}{2012}).

\bibitem{Clauset-Shalizi-Newman-2009-SIAMR}
\bibinfo{author}{Clauset, A.}, \bibinfo{author}{Shalizi, C.~R.} \&
  \bibinfo{author}{Newman, M. E.~J.}
\newblock \bibinfo{title}{{Power-law distributions in empirical data}}.
\newblock \emph{\bibinfo{journal}{SIAM Rev.}} \textbf{\bibinfo{volume}{51}},
  \bibinfo{pages}{661--703} (\bibinfo{year}{2009}).

\bibitem{Arianos-Carbone-2007-PA}
\bibinfo{author}{Arianos, S.} \& \bibinfo{author}{Carbone, A.}
\newblock \bibinfo{title}{{Detrending moving average algorithm: A closed-form
  approximation of the scaling law}}.
\newblock \emph{\bibinfo{journal}{Physica A}} \textbf{\bibinfo{volume}{382}},
  \bibinfo{pages}{9--15} (\bibinfo{year}{2007}).

\bibitem{Jiang-Xie-Zhou-2014-PA}
\bibinfo{author}{Jiang, Z.-Q.}, \bibinfo{author}{Xie, W.-J.} \&
  \bibinfo{author}{Zhou, W.-X.}
\newblock \bibinfo{title}{{Testing the weak-form efficiency of the WTI crude
  oil futures market}}.
\newblock \emph{\bibinfo{journal}{Physica A}} \textbf{\bibinfo{volume}{405}},
  \bibinfo{pages}{235--244} (\bibinfo{year}{2014}).

\bibitem{Meneveau-Sreenivasan-1987-PRL}
\bibinfo{author}{Meneveau, C.} \& \bibinfo{author}{Sreenivasan, K.~R.}
\newblock \bibinfo{title}{{Simple multifractal cascade model for fully
  developed turbulence}}.
\newblock \emph{\bibinfo{journal}{Phys. Rev. Lett.}}
  \textbf{\bibinfo{volume}{59}}, \bibinfo{pages}{1424--1427}
  (\bibinfo{year}{1987}).

\bibitem{Hawkes-1971-Bm}
\bibinfo{author}{Hawkes, A.}
\newblock \bibinfo{title}{{Spectra of some self-exciting and mutually exciting
  point processes}}.
\newblock \emph{\bibinfo{journal}{Biometrika}} \textbf{\bibinfo{volume}{58}},
  \bibinfo{pages}{83--90} (\bibinfo{year}{1971}).

\bibitem{Filimonov-Sornette-2012-PRE}
\bibinfo{author}{Filimonov, V.} \& \bibinfo{author}{Sornette, D.}
\newblock \bibinfo{title}{{Quantifying reflexivity in financial markets: Toward
  a prediction of flash crashes}}.
\newblock \emph{\bibinfo{journal}{Phys. Rev. E}} \textbf{\bibinfo{volume}{85}},
  \bibinfo{pages}{056108} (\bibinfo{year}{2012}).

\bibitem{Jiang-Xie-Li-Zhou-Sornette-2016-JSM}
\bibinfo{author}{Jiang, Z.-Q.}, \bibinfo{author}{Xie, W.-J.},
  \bibinfo{author}{Li, M.-X.}, \bibinfo{author}{Zhou, W.-X.} \&
  \bibinfo{author}{Sornette, D.}
\newblock \bibinfo{title}{{Two-state Markov-chain Poisson nature of individual
  cellphone call statistics}}.
\newblock \emph{\bibinfo{journal}{J. Stat. Mech.-Theory Exp.}}
  \textbf{\bibinfo{volume}{2016}}, \bibinfo{pages}{073210}
  (\bibinfo{year}{2016}).

\end{thebibliography}

\end{document}